\documentclass{article}

\usepackage{arxiv}

\usepackage{algorithmicx}
\algdef{SE}[DOWHILE]{Do}{doWhile}{\algorithmicdo}[1]{\algorithmicwhile\ #1}%
\usepackage{verbatim}
\usepackage[export]{adjustbox}
\usepackage{algpseudocode}
\makeatletter
\def\BState{\State\hskip-\ALG@thistlm}
\makeatother
\usepackage{algorithm}
\usepackage{amssymb}
\usepackage{color, colortbl}
\usepackage{float}
\newcolumntype{y}{>{\columncolor{yellow}}c}
\usepackage{hyperref}
\usepackage[english]{babel}
\usepackage{tikz}
\usetikzlibrary{positioning,shapes,shadows,arrows}
\usepackage{amsfonts}
\usepackage{amsthm}
\usepackage{amsmath}
\usepackage{mathrsfs}
\usepackage{booktabs}
\usepackage{mathtools}
\usepackage{amssymb,epsfig,multirow}
\usepackage{amsbsy}
\usepackage{graphicx}
\usepackage{geometry}
\usepackage{url}
\usepackage{multicol}
\usepackage{float}
\usepackage{color}
\usepackage{enumerate}

\usepackage[pdf]{graphviz}
\usepackage{tikz}
\usetikzlibrary{arrows.meta}
\usepackage{tkz-euclide}
\usepackage{tkz-graph}
\usetikzlibrary{graphs,quotes,arrows.meta}
\usepackage[utf8]{inputenc} 
\usepackage[T1]{fontenc}    
\usepackage{hyperref}       
\usepackage{url}            
\usepackage{booktabs}       
\usepackage{amsfonts}       
\usepackage{nicefrac}       
\usepackage{microtype}      
\usepackage{lipsum}		
\usepackage{graphicx}
\usepackage{doi}

\usepackage{algorithmicx}
\usepackage{times}
\newtheorem{theorem}{Theorem}
\theoremstyle{plain}

\theoremstyle{definition}
\theoremstyle{remark}

\newtheorem{prop}[theorem]{Proposition}
\newtheorem{problem}{Problem}

\newtheorem{exam}{Example}[section]

\newtheorem{defn}{Definition}[section]
\newcommand{\Real}{\mathbb{R}}
\newcommand{\C}{{\mathcal{C}}}
\newcommand{\Q}{{\mathcal{U}}}
\newcommand{\E}{{\mathcal{E}}}
\newcommand{\A}{{\mathcal{A}}}

\newcommand{\B}{{\mathcal{B}}}

\newcommand{\Ss}{{\mathcal{S}}}

\renewcommand{\L}{\mathcal{L}}

\title{Dynamic Programming based Local Search approaches for Multi-Agent Path Finding problems on Directed Graphs}

\title{\LARGE \bf {Dynamic Programming based Local Search approaches for Multi-Agent Path Finding problems on Directed Graphs}}

\author{I. Saccani$^{a}$, S. Ardizzoni$^{a}$,  L. Consolini$^{a}$,  M. Locatelli$^{a}$  \\
\vspace{0.3cm}
       {  \footnotesize $^{a}$\textit{Dipartimento di Ingegneria e Architettura, Universit\`a di Parma, Parco Area delle Scienze, 181/A, Parma, Italy}}
}

\date{}

\renewcommand{\headeright}{}
\renewcommand{\undertitle}{}
\renewcommand{\shorttitle}{}

\hypersetup{
pdftitle={Dynamic Programming based Local Search approaches for Multi-Agent Path Finding problems on Directed Graphs},
pdfsubject={cs.MA},
pdfauthor={I. Saccani, S. Ardizzoni,  L. Consolini,  M. Locatelli}
}

\begin{document}
\maketitle

\begin{abstract}
Among sub-optimal Multi-Agent Path Finding (MAPF) solvers, rule-based algorithms are particularly appealing since they are complete. Even in crowded scenarios, they allow finding a feasible solution that brings each agent to its target, preventing deadlock situations. However, generally, rule-based algorithms provide much longer solutions than the shortest one. 
The main contribution of this paper is introducing a new local search procedure for improving a known feasible solution. We start from a feasible sub-optimal solution, and perform a local search in a neighborhood of this solution. If we are able to find a shorter solution, we repeat this procedure until the solution cannot be shortened anymore.
At the end, we obtain a solution that is still sub-optimal, but generally of much better quality than the initial one.
We propose two different local search policies. In the first, we explore all paths in which the agents positions remain in a neighborhood of the corresponding positions of the reference solution. In the second, we set an upper limit to the number of agents that can change their path with respect
to the reference solution. These two different policies can also be alternated. We explore the neighborhoods by dynamic programming. The fact that our search is \emph{local} is fundamental in terms of time complexity. Indeed, if the dynamic programming approach is applied to the full MAPF problem, the number of explored states grows exponentially with the number of agents. Instead, the introduction of a locality constraint allows exploring the neghborhoods
in a time that grows polynomially with respect to the number of agents.
\end{abstract}

\section{INTRODUCTION}
We consider the Multi-Agent Path Finding (MAPF) problem. This problem is defined on a directed graph. The nodes represent the positions occupied by a set of agents. The arcs represent the allowed motions between nodes.
 Each agent occupies a different node and can move to free nodes (i.e., nodes not occupied by other agents).
The MAPF problem consists in computing a sequence of movements that repositions all agents to assigned target nodes, avoiding collisions. 
The main motivation comes from managing fleets of automated guided vehicles (AGVs). AGVs move items between locations in a warehouse. Each AGV follows predefined paths, that connect the locations where items are stored or processed. We represent the warehouse layout with a directed graph.
The nodes represent positions where items are picked up and delivered,
together with additional locations used for routing. The directed arcs represent the precomputed paths that connect these locations. If various AGVs move in a small scenario, each AGV represents an obstacle for the others. In some cases, the fleet can reach a deadlock situation, in which some vehicles are unable to reach their target. Hence, it is important to find a feasible solution to MAPF, even in crowded configurations.

{\bf Literature review.}
Various works address the problem of finding the optimal solution of MAPF (i.e., the solution with the minimum number of moves). For instance, Conflict Based Search (CBS) is a two-level algorithm that uses a search tree, based on 
conflicts between individual agents (see~\cite{sharon2015}).
However, finding the optimal solution of MAPF is NP-hard (see~\cite{yu2013}), and computational time grows exponentially with the number of agents. Therefore, typically, optimal solvers are only applied when the number of agents is relatively small. Conversely, sub-optimal solvers are usually employed when the number of agents is large. In such cases, the aim is to quickly find a feasible solution. That is, a sequence of motions that positions each agent on the assigned target configuration. In general, the provided solution is not the optimal one. Among sub-optimal solvers, we can distinguish search-based and rule-based approaches.
Search-based solvers aim to provide a high-quality solution but are not complete (i.e., they are not always able to return a feasible solution). A prominent example is Hierarchical Cooperative A$^*$ (HCA$^*$) \cite{silver2021}, in which agents are planned one at a time, according to some predefined order.
Instead, rule-based approaches include specific movement rules for different scenarios. They favor completeness at low computational cost over solution quality. One of the first important results for rule-based algorithms is from Kornhauser's thesis \cite{kornhauser1984}, which presents a rule-based procedure to solve MAPF (or to establish that MAPF has no feasible solution). Kornhauser shows that the solutions found with his method have cubic length complexity, with respect to the number of nodes and agents. However, the algorithm proposed by Kornhauser is quite complex and, to our knowledge, has never been fully implemented. Two relevant  rule-based algorithms are TASS \cite{khorshid2021} and \textit{Push and Rotate} \cite{wilde2014} \cite{alotaibi2016}. TASS is a tree-based agent swapping strategy that is complete on every tree, while \textit{Push and Rotate} solves every MAPF instance on graphs that contain at least two unoccupied vertices. Reference~\cite{krontiris2021} presents a method that converts the graph into a tree (as in~\cite{goraly2010}) and solves the resulting problem with TASS. Recently, reference~\cite{ardizzoni2024} proposed the \textit{Leaves Procedure}, simple
and easy to implement, which finds solutions to MAPF on trees with a lower length complexity, compared to both TASS and Kornhauser's algorithm. Rule-based algorithms are also used for directed graphs with at least two unoccupied nodes. In particular, reference~\cite{ardizzoni2022} presents the diSC algorithm, which solves any MAPF instance on strongly connected digraphs (i.e., directed graphs
in which it is possible to reach any node starting from any
other node).
Another relevant reference is~\cite{botea2018}, which solves MAPF on the specific class of biconnected digraphs (i.e., strongly connected digraphs where the undirected graphs obtained by ignoring the edge orientations have no cutting vertices). The proposed diBOX algorithm has polynomial complexity with respect to the number of nodes.
Among sub-optimal MAPF solvers, rule-based algorithms are particularly appealing, since they are complete. Even in crowded scenarios, they allow finding a feasible solution that brings each agent to its target, preventing deadlock situations. However, generally, rule-based algorithms provide solutions that are much longer than the shortest one. This is a crucial limitation in industrial applications. 
For this reason, a third class of algorithms, combining optimal and sub-optimal solvers, the class of anytime MAPF algorithms, is of particular relevance. Approaches in this class
    aim at first detecting quickly a feasible solution, and then at improving it through suitable procedures.
    Among them, those based on Large Neighborhood Search (LNS) \cite{shaw1998} are particularly well-known. LNS is a popular local search technique to improve the solution quality for combinatorial optimization problems. Starting from a given solution, it removes part of the solution, called a neighborhood, and treats the remaining part of the solution as fixed. Then, it repairs the solution and replaces the old solution if the repaired solution is better. This procedure is repeated until some stopping criterion is met. For instance, MAPF-LNS \cite{li2021} is an anytime MAPF framework that improves the quality of a solution obtained from a MAPF algorithm over time by replanning subgroups of agents using LNS. In \cite{li2021}, the number of agents to be replanned is a predefined parameter $N$, and three different rules to select them
    are proposed, one based on agent paths, another based on graph topology, and, finally, a random one. We also mention \cite{li2022}, where algorithm LNS2 is introduced. Such algorithm 
    starts from an infeasible solution and repeatedly replans subsets of agents to find plans with fewer conflicts, eventually converging to a conflict-free plan.
	    \newline\newline\noindent
{\bf Statement of contribution.} 
The aim of this paper is to propose methods to shorten a given feasible sub-optimal solution of a MAPF instance on a directed graph, expanding the results of our conference paper~\cite{ardizzoni231}. In that reference, we propose an iterative local search approach where neighborhoods are explored through dynamic programming. Namely, we start from a feasible sub-optimal solution, define a neighborhood, and search for a better solution within the neighborhood by dynamic programming. If a better solution is found, we repeat the whole procedure, otherwise we stop and the current solution is a local minimizer with respect to the given neighborhood.
Given a distance parameter $r$ and a function measuring the distance from the reference solution, the neighborhood is the set of solutions
whose distance from the reference solution is not larger than $r$. Different ways to measure the distance lead to different approaches. As explained in~\cite{ardizzoni231}, the fact that our search is \emph{local} is fundamental for reducing the time complexity of the proposed algorithms. Indeed, in principle, it is possible to solve to (global) optimality the MAPF problem by dynamic programming. However, the number of explored states grows exponentially with the number of agents, so we cannot apply standard dynamic programming to problems with many agents.  The introduction of a locality constraint allows solving the (local) problem through dynamic programming in a time that grows only polynomially with respect to the number of agents (see Theorem~\ref{thm_complx}).
The following are the main new contributions with respect to our conference paper~\cite{ardizzoni231}.

\begin{enumerate}
\item  We present a new local search method, where the distance from a reference feasible solution is measured in terms of number of agents that change their paths with respect to such solution.
This is related but different with respect to what is done in MAPF-LNS~\cite{li2021}. In MAPF-LNS, a fixed subset of agents with cardinality $N$ is selected in advance and only the agents in that subset are allowed to move. Instead, in our approach,  we fix the number $N$ of agents which can change their positions with respect to the reference solution, but we do not fix in advance such $N$ agents, i.e., we consider all possible subsets of $N$ agents. Note that, with respect to LNS, we employ lower values for $N$, to avoid the exploration of too large neighborhoods.
\item We propose a more general local search method in which we alternate different policies, namely the one presented in reference \cite{ardizzoni231}  and the new one introduced in this paper. This alternating method allows for improvements in the solutions, as it will be shown in the simulations.
\item  As already mentioned, for all the proposed approaches we prove polynomial time complexity when the distance parameter $r$ is fixed.
\item We extend our numerical simulations, comparing the results obtained through the different policies.
  \end{enumerate}

  \subsection{Notation}
\label{sec:notation}
A directed graph is a pair $G=(V,E)$ where $V$ is a set of
nodes
and $E \subset \{(x,y) \in V^2 \mid x \neq y\}$ is a set of directed arcs.
A path $p$ on $G$ is a sequence of adjacent nodes of $V$
(i.e., $p= \sigma_{1}\cdots \sigma_{m}$, with $(\forall i \in \{1, \ldots, m\})\
(\sigma_i,\sigma_{i+1}) \in E$).
An alphabet $\Sigma = \{\sigma_1,\ldots,\sigma_n\}$ is a set of symbols.
A word is any finite sequence of symbols.
The set of all words over $\Sigma$ is $\Sigma^*$, which also contains
the empty word $\epsilon$.
Given a word $w \in \Sigma^*$, $|w|$ denotes its length.
Given $s,t \in \Sigma^*$, the word obtained by writing $t$ after $s$ is
the concatenation of $s$ and $t$, denoted by $st \in \Sigma^*$;
we call $t$ a suffix of $st$ and $s$ a prefix of $st$.
We use prefixes and suffixes to represent portions of paths.
A Finite State Automaton (FSA) is a quintuple $(\Sigma,X,x_0,\delta,X_M)$, where $\Sigma$ is a finite set of symbols, $X$ is a set of states, $x_0 \in S$ is the initial state, $\delta: X \times \Sigma \to X$ is the transition function and $X_M \subset S$ is a set of final states. Function $\delta$ can be partially defined, that is, not defined for all elements of the domain. Given $x \in X$, $\sigma \in \Sigma$, if $\delta(x,\sigma)$ is defined, we use notation $\delta(x,\sigma)!$. Function $\delta$ can be extended to a function $\delta: X \times \Sigma^* \to X$ by setting $\delta(x,s \sigma)=\delta(\delta(x,s),\sigma)$ and $\delta(x,\epsilon)=x$.

\section{Problem definition}
\label{sec:prob_def}
Let $G=(V, E)$ be a directed graph, with vertex set $V$ and edge set $E$. We identify each agent with a unique label, and the ordered set $P$ contains these labels. A \textit{configuration} is a function  $\A:P \rightarrow V$ that assigns the occupied vertex to each agent. A configuration is \textit{valid} if it is injective (i.e., each vertex is occupied by at most one agent). Set $\C \subset \{P \to V\}$ represents all valid configurations. We represent a configuration $\A \in \C$ also by an ordered set of vertices $(v_1,v_2,\ldots,v_r)$, where $v_i$, $i=1,\ldots,r$ is the vertex occupied by the $i$-th agent. Time is assumed to be discretized. At every time step, each agent occupies one vertex and executes a single action. There are two types of actions: \emph{wait} and \emph{move}. We denote the wait action by $\iota$. An agent that executes this action remains in its current vertex for another time step. We denote a move action by $u \rightarrow v$.  In this case, the agent moves from its current vertex $u$ to an adjacent vertex $v$  (i.e., $(u,v) \in E$). Therefore, the set of all possible actions for a single agent is
	\[ \bar{E} = E \cup \{\iota\}.\]
Function $\rho: \C \times \bar{E} \rightarrow \C$ is a partially defined transition function such that $\A' =\rho(\A,u \rightarrow v)$  is the configuration obtained by moving an agent from $u$ to $v$:
\begin{equation} 
	\label{c}
	\A'(q):=  \Bigg\{
	\begin{array}{ll}       	
		v, & \text{if } \A(q)  = u ;\\		
		\A(q), & \text{otherwise } .\\
	\end{array}
\end{equation}
Notation $\rho(\A,u \rightarrow v)!$ means that the function is well-defined. In other words $\rho(\A, u \rightarrow v)!$ if and only if $(u,v)\in E$ and $\A' \in \C$.  Moreover, $(\forall \A \in \C)\; \rho(\A,\iota)!$ and $\rho(\A,\iota)=\A$.
Since multiple agents can move at the same time step, an action of the whole fleet  is an element $a=(a_1,\dots ,a_{|P|})$ of $\E=\bar{E}^{|P|}$,  where $a_i$ is the action of agent $i$. We can extend function $\rho:\C \times \bar{E} \to \C$ to $\rho: \C \times \E  \to \C$, by setting  $\A'=\rho(\A,  a) $ equal to the configuration obtained by moving agent $i$ along edge $a_i$ (or by not moving the agent if $a_i=\iota$). In this case, $(\forall a \in \E, \A \in \C)$  $\rho(\A,  a) !$ if and only if the following conditions hold:
\begin{enumerate}
 	\item $\A' \in \C$: two or more agents cannot occupy the same vertex at the same time step;
     \item $\forall i=1,\dots, |P|$, if $a_i=(u,v)$, then $\not \exists j \in \{1,\dots, |P|\}$ such that $a_j=(v,u)$: two agents cannot swap locations in a single time step.
\end{enumerate} 	
We represent plans as ordered sequences of actions. It is convenient to view the elements of $\E$ as the symbols of a language.	We denote by $\E^*$ the Kleene star of $\E$, that is the set of ordered sequences of elements of $\E$ with arbitrary length, together with the empty string $\epsilon$:
\[
\E^*=\bigcup_{i=1}^\infty \E^i \cup \{\epsilon\}.
\]
We extend function $\rho:\C \times \E \to \C$ to $\rho: \C \times \E^*  \to \C$ according to the following rules
\begin{enumerate}
	\item $(\forall s \in \E^*, e \in \E, \A \in \C)$  $\rho(\A,  se) !$ if and only if $\rho(\A, s)!$ and $\rho(\rho(\A, s), e)!$
	\item if $\rho(\A, se)!$, then $\rho(\A, se)=\rho(\rho(\A, s),e)$.
  \end{enumerate}
Note that $\epsilon$ is the trivial plan that keeps all agents at their positions. In some cases, it is convenient to represent the individual plan of each agent, defining a plan $f$ as an element of $((\bar E)^*)^{|P|}$. In this way,  $f \in \E^*$ is a $|P|$-tuple $f=(f^1, \cdots, f^{|P|})$, where $f^i = \{a_i^1,\cdots, a_i^{|f|}\}$ represents the ordered sequence of the actions of agent $i$. For all $i=1,\cdots, |P|$, we call $f^i$ the \textit{path plan of agent} $i$. We denote by $\L$ the set of plans such that $\rho(\A,  f) $ is well-defined:
\[\L = \{ f \in \E^* : \, \rho(\A,  f)!\}. \]

The problem of detecting a feasible solution is the following:
\begin{problem}{(\textbf{Feasibility MAPF problem}).}
  \label{prob_feas}
	Given a digraph $G = (V,E)$, an agent set $P$, an initial valid configuration $\A$, and a final valid configuration $\A^t$, find a plan $f\in  \L$ such that $\A^t = \rho(\A,f)$.
      \end{problem}

It is natural to represent the agents behaviour by a FSA $(\E,\C,\A,\rho,\{A^t\})$, where the alphabet $\E$ is the set of actions, the state set $\C$ consists in valid configurations, $\A$ is the initial state, $\rho$ is the transition function, and $\A^t$ is the final state. In this way, Problem~\ref{prob_feas} corresponds to a reachability task for the FSA.

For a feasible plan $f$, we define $|f|$ as the length of plan $f$ (i.e., the number of time steps needed for all agents to reach the final configuration, through plan $f$). Furthermore, given $k \in  {\mathbb{N}} $, we denote by $f_k$ the $k$-th prefix of $f$ (that is, the prefix of $f$ of length $k$, made up of the first $k$ actions of $f$). Note that $|f_k|=k$. Also note that $|f|$ corresponds to the cost function usually called \textit{Makespan}. Given a MAPF instance, with initial configuration $\A$ and final configuration $\A^t$, the \emph{optimization MAPF problem} aims at finding a feasible plan $f$ of minimal length $|f|$. 
\begin{problem}{(\textbf{Optimization MAPF problem}).}
Let $\A$ and $\A^t$ be valid configurations on a digraph $G$, solve
\begin{equation} 
	\label{op}
	\centering 
	\begin{array}{ll}       	
		
	\min{|f|} & \\
	
	      & \\
		
	\quad 	\text{s.t.} \;\; \A^t = \rho(\A,f) &  \\
	
	  & \\
	
	\quad   \quad  \quad f \in \L. & \\
	
	\end{array}
\end{equation}
\end{problem}
Other cost functions have also been used in the literature. \textit{Sum-of-costs}, for example, is the sum of the number of time steps that each agent employs to reach its target, without leaving it again. Unfortunately, finding the optimal solution (i.e., the minimal \textit{Makespan} or \textit{sum-of-costs}) is NP-hard \cite{yu2013}. Intuitively, this happens since the set of possible plans grows exponentially with respect to the number of agents and nodes.
In this paper, we propose an approach to find a good quality sub-optimal solution in polynomial time. Roughly speaking, the method we propose is a local search outlined in what follows:
\begin{itemize}
\item Using a rule-based algorithm, we find a plan $f$ that solves the feasibility MAPF problem.
\item We define a neighborhood of plan $f$, whose size grows polynomially with respect to the number of agents and nodes. In this neighborhood, we look for a feasible solution $\hat f$, shorter than $f$. 
\item If we can find such a solution, we set $f=\hat f$ and reiterate the local search. Otherwise, we stop and return the current plan.
\end{itemize}
This algorithm runs in polynomial time with respect to the number of agents and nodes. Indeed, under mild assumptions, a rule-based algorithm provides a feasible solution in polynomial time, and this solution has polynomial length. At each iteration, the length is decreased by at least one, so that the number of iterations is polynomially bounded. Finally, we will define neighborhoods with polynomial cardinality with respect to the number of agents and nodes. Thus, they can be explored in polynomial time. At the end of the procedure, we obtain, in polynomial time, a solution to the MAPF problem that is locally optimal, with respect to the employed neighborhood, but not necessarily globally optimal. 

\section{Measuring the distance between plans}
\label{sec:distances}
The key point of this paper is to define suitable neighborhoods of a reference plan $f$, whose cardinality grows polynomially with respect to the number of agents and nodes. As a first step, we show possible ways to measure the distance between a generic plan $g$ and a reference plan $f$. We introduce two families of distances:
\begin{itemize}
\item
  In the first family (path distances) we take into account the graph distance between the nodes occupied by the agents in $f$ and $g$. 
\item In the second family (agent distances) we count the number of agents in $f$ and $g$ that have different individual plans.
  \end{itemize}
In the following, we define these two families in more detail.

\subsection{Path distances}
\label{subsec:pathsdistances}

In this section, we focus on distances based on the length of the shortest paths connecting the vertices of the graph. 
\begin{defn}
Let $G=(V, E)$ be a digraph and $P$ be a set of agents. We define the distance of vertex $u$ from vertex $v$ as the length of the shortest path on $G$ from $v$ to $u$:
\begin{equation}
 \hat d: V \times V \rightarrow  {\mathbb{N}} \quad \quad \hat d(u,v) = \ell(\pi_{vu}),
\label{eq:dist_vert}
\end{equation}
where $\pi_{vu}$ is the shortest path in $G$ from $v$ to $u$ and $\ell(\pi_{vu})$ is the length of that path, defined as the number of edges of $\pi_{vu}$. 
\end{defn}
Note that $\hat d$ is not symmetrical, since $\pi_{uv}$ and $\pi_{vu}$ can be different. Next, we define the distance of configuration $\A^1$ from configuration $\A^2$ as the sum of the distances between the vertices that each agent occupies in the two configurations:
\[ \bar d: \C \times \C \rightarrow  {\mathbb{N}} \quad \quad \bar{d} (\A^1,\A^2) = \sum_{p \in P} \hat d(\A^1(p),\A^2(p)).\]
Finally, we define the asymmetrical distance between two plans in $\L$. To this end, we associate to each plan $f$ a function $\psi_f: \mathbb{N} \to \C$, so that $\psi_f(k)$ is the configuration corresponding to plan $f$ at step $k$. If $k > |f|$, that is the step $k$ is larger than the length of $f$, $\psi_f(k)$ is the last configuration:
\[\psi_f(k):= \bigg\{ \begin{array}{ll}\rho(\A,f_k), & k < |f|, \\
\rho(\A,f), & k \geq |f|.
  \end{array}
\]

Note that $\psi_f(k)$, the configuration at step $k$, depends on $f_k$, the $k$-th prefix of $f$. We define the distance of plan $f$ from plan $g$ with respect to the associated functions $\psi_f,\psi_g$.
We consider the following distances:
\begin{enumerate}
\item $\infty$-distance:
\begin{equation}
  d_{\infty}(f,g) := \max_{1 \leq k \leq \min \{|f|,|g|\}}\; \bar{d} (\psi_f(k), \psi_g(k)) ;
  \label{eq:inf_dist}
\end{equation}
\item $1$-distance: 
\begin{equation}
  d_{1}(f,g) := \sum_{k =1}^{ \min \{|f|,|g|\}}\; \bar{d} (\psi_f(k), \psi_g(k)) ;
  \label{eq:one_dist}
\end{equation}
\item max-min distance:
\begin{equation}
\label{eq:max_min}
d_{\infty}^*(f,g) :=\max_{k \in {\mathbb{N}}}\; \min_{h \in {\mathbb{N}}}\bar{d} (\psi_f(k), \psi_g(h)) ;
\end{equation}

\item sum-min distance:
\begin{equation}
\label{eq:sum_min}
d_{1}^*(f,g):= \sum_{k =1}^{ \min \{|f|,|g|\}} \min_{h \in {\mathbb{N}}} \bar{d} (\psi_f(k), \psi_g(h)).
\end{equation}
\end{enumerate}

\begin{figure}[!ht]
\centering
\includegraphics[width=0.45\textwidth]{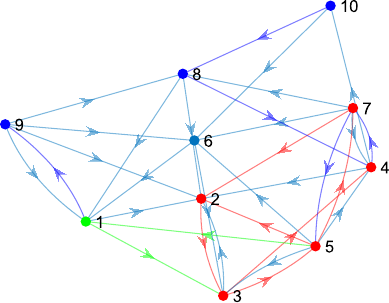}\par
\caption{Plan $f_0$ on graph with 10 nodes.}
\label{fig:graph10}
\end{figure}

\begin{figure}[!ht]
\centering
\includegraphics[width=0.45\textwidth]{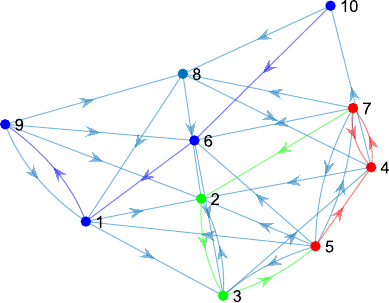}\par
\caption{Plan $g$ on graph with 10 nodes.}
\label{fig:graph10ls}
\end{figure}

Namely, the $\infty$-distance of plans $f$ and $g$ corresponds to the maximum, with respect to time-step $k$, of the distance between the corresponding configurations at $k$. The $1$-distance corresponds to the sum, with respect to time-step $k$, of the distances between the corresponding configurations at $k$. The max-min distance (respectively, the sum-min distance) corresponds to the maximum (respectively, the sum) with respect to $k$, of the distance of the configuration that plan $f$ reaches at step $k$ with respect to the set of all configurations encountered by plan $g$. It is easy to see that, for each couple of plans $f$, $g$, the $1$-distance is the largest of the four, while the max-min distance is the smallest.
\begin{exam}
  Figure~\ref{fig:graph10} shows a directed graph with 10 nodes. Tables~\ref{tab:f0} and \ref{tab:g} define two plans $f_0$ and $g$, depicted also in Figures~\ref{fig:graph10} and~\ref{fig:graph10ls}. All configurations from $k=0$ to $k=2$ are the same in the two plans. All configurations in $g$, with the exception of the one for step $k=3$, are also present in $f_0$. At $k=3$, plan $g$ has a configuration $(7,6,5)$. The configuration in $f_0$ that is closer to $(7, 6, 5)$ is the one at $k=3$, that is  $(7, 10, 5)$. Agent $2$ is the only agent that occupies different nodes ($6$ and $10$) in the two configurations. The shortest path between node $6$ and node $10$ has $3$ edges. Hence, both the sum-min distance and the max-min distance between the two plans are $3$. That is, $d_\infty^*(g,f_0)=3$, $d_1^*(g,f_0)=3$. Finally, the configurations of plan $g$ at $k= 4$ and $k=5$ have a distance from the corresponding configurations of plan $f_0$ both equal $8$. Therefore, the $\infty$-distance is equal to $8$ and the $1$-distance is equal to $3+8+8=19$. That is, $d_\infty(g,f_0)=8$, $d_1(g,f_0)=19$.
\end{exam}

\begin{table}[!ht]
\begin{center}
  \begin{tabular}{| l ||  c | c | c | y | c | c | c | c | c | c | c | c | c | c | r | }
    \hline
    k & 0 & 1 & 2 & 3 & 4 & 5 & 6 & 7 & 8 & 9 & 10 & 11 & 12 & 13 \\ \hline \hline
    \textbf{1} & 5 & 7 & 7 & 7 & 2 & 3 & 5 & 2 & 2 & 2 & 2 & 3 & 4 & 4 \\ \hline
    \textbf{2} & 10 & 10 & 10 & 10 & 10 & 10 & 10 & 8 & 4 & 7 & 5 & 1 & 1 & 9 \\ \hline
    \textbf{3} & 7 & 2 & 3 & 5 & 1 & 1 & 1 & 1 & 3 & 3 & 3 & 5 & 5 & 5 \\
    \hline
  \end{tabular}
\end{center}
\caption{Configurations of plan $f_0$, the $k$-th column contains the nodes occupied by the agents at step $k$}

   \label{tab:f0}
\end{table}

\begin{table}[!ht]
\begin{center}
  \begin{tabular}{| l ||  c | c | c | y | c | c | r | }
    \hline
    k & 0 & 1 & 2 & 3 & 4 & 5 \\ \hline \hline
    \textbf{1} & 5 & 7 & 7 & 7 & 4 & 4 \\ \hline
    \textbf{2} & 10 & 10 & 10 & 6 & 1 & 9 \\ \hline
    \textbf{3} & 7 & 2 & 3 & 5 & 5 & 5 \\
    \hline
  \end{tabular}
\end{center}
\caption{Configurations of plan $g$.}
   \label{tab:g}
\end{table}

\subsection{Agent Distances}
\label{sec:a_d}
Alternatevely, we can define the distance between two plans as the number of agents that occupy different positions in the two plans. First, we define the agent distance between two configurations. Let $U: \C \times \C \rightarrow {\mathbb{P}(P)}$ be a function that associates to each pair of configurations the subset of agents that are on different nodes in the two configurations:
\begin{equation}
	\label{eq:Q}
U: \C \times \C \rightarrow  {\mathbb{P}(P)} \quad \quad U(\A_1,\A_2) := \Bigl\{p \in P \; : \;\A_1(p) \neq \A_2(p)\Bigl\},	
\end{equation}
The distance between two configurations  $\A_1$ and $\A_2$ is the cardinality of  $U(\A_1, \A_2)$:
\begin{equation}
	\label{eq:dist_ag_conf}
	\tilde{d}(\A_1, \A_2) := |	U(\A_1, \A_2)|.
	\end{equation}
Similarly, we define a new function $\Q$ which associates to each pair of plans $f$ and $g$ the subset of agents that have different positions in at least one time step $ k \in \{1,\ldots, max(|f|, |g|)\}$
\begin{equation}
	\label{eq:Q_fg}
\Q: \L \times \L \rightarrow {\mathbb{P}(P)}
\quad \Q(f, g) :=  \Bigl\{p \in P \; : \;\exists k \in \{1,\ldots, max(|f|, |g|)\}, \;(\psi_f(k))(p) \neq (\psi_g(k))(p)\Bigl\}.
\end{equation}

\noindent
Note that:
\begin{equation}
	\Q(f, g) =  \bigcup_{k=1}^{max(|f|, |g|)} 	U(\psi_f(k), \psi_g(k)),
	\label{eq:Q_u}
\end{equation}
\noindent
or, equivalently:

\begin{equation}
	\Q(f, g) =  \Bigl\{p \in P \; : \;f^p \neq g^p\Bigl\}.
	\label{eq:Q_u2}
\end{equation}
We define a new agent distance between two plans, called \textit{u-agents distance}, as the cardinality of $\Q(f,g)$:
\begin{equation}
	\label{eq:dist_ag}
	d_U(f,g) := \Big| \Q(f,g) \Big|.
\end{equation}
\noindent

Namely, this distance represents the number of agents that, in the two plans, occupy different positions in at least one time-step or, equivalently, the number of agents that have a different \textit{path plan} in $f$ and $g$.

Alternatively, we can define another agent distance $d_m$ called \textit{max-agents distance} between two plans $f$ and $g$ as 
the maximum of all the distances between the pairs of configurations crossed by the two plans as the time step varies from $1$ to  $ max(|f|, |g|)$

\begin{equation}
\label{eq:dist_max_ag}
	d_m(f,g) := \max_{k \in \{1,\ldots, max(|f|, |g|)\},}\; 	\tilde{d}(\psi_f(k), \psi_g(k)),
\end{equation}
where $\tilde{d}(\psi_f(k), \psi_g(k))$ is the distance (\ref{eq:dist_ag_conf}) between configurations.
\noindent
The following relationship exists between the two distances defined above.
\begin{prop}
\label{prop:dist}
Let $f,g \in \L$ be two plans. It holds that:
\[
 d_m(f,g) \leq d_U(f,g).
\]

\end{prop}

\begin{proof}
From (\ref{eq:Q_u}) it follows that:
\[
\forall k \in {\mathbb{N}}, \; \; |U(\psi_f(k), \psi_g(k))| \leq |\Q(f,g)|,
\]
so that:
\[
\forall k \in {\mathbb{N}}, \; \; \tilde{d}(\psi_f(k), \psi_g(k)) \leq d_U(f,g),
\]
and, consequently:
\[
\max_{k \in {\mathbb{N}}}\; \tilde{d}(\psi_f(k), \psi_g(k)) \leq d_U(f,g).
\]

\end{proof}

\begin{figure}[!ht]
\centering
\includegraphics[width=0.45\textwidth]{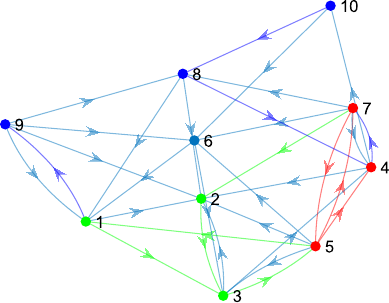}\par
\caption{Plan $h$ on graph with 10 nodes.}
\label{fig:graph10tw}
\end{figure}

\begin{exam}
\label{ex:agents}
Consider plan $f_0$ in Table~\ref{tab:f0_2}, and plan $h$ in Table~\ref{tab:h}, depicted also in Figures~\ref{fig:graph10} and~\ref{fig:graph10tw}. The u-agents distance and the max-agents distance between the two plans are $1$ because only the first agent has different path plans in $f_0$ and $h$.
\end{exam}
\begin{table}[!ht]
\begin{center}
  \begin{tabular}{| l ||  c | c | c |  c | c | c | c | c | c | c | c | c | c | c | r | }
    \hline
    k & 0 & 1 & 2 & 3 & 4 & 5 & 6 & 7 & 8 & 9 & 10 & 11 & 12 & 13 \\ \hline \hline
    \rowcolor{yellow}
    \textbf{1} & 5 & 7 & 7 & 7 & 2 & 3 & 5 & 2 & 2 & 2 & 2 & 3 & 4 & 4 \\ \hline
    \textbf{2} & 10 & 10 & 10 & 10 & 10 & 10 & 10 & 8 & 4 & 7 & 5 & 1 & 1 & 9 \\ \hline
    \textbf{3} & 7 & 2 & 3 & 5 & 1 & 1 & 1 & 1 & 3 & 3 & 3 & 5 & 5 & 5 \\
    \hline
  \end{tabular}
\end{center}
\caption{Initial plan $f_0$.}

   \label{tab:f0_2}
\end{table}
\begin{table}[!ht]
\begin{center}
  \begin{tabular}{| l ||  c | c | c |  c | c | c | c | c | c | c | c | c | c | c | r | }
    \hline
    k & 0 & 1 & 2 & 3 & 4 & 5 & 6 & 7 & 8 & 9 & 10 & 11 & 12 & 13 \\ \hline \hline
        \rowcolor{yellow}
    \textbf{1} & 5 & 7 & 7 & 7 & 5 & 5 & 5 & 5 & 5 & 4 & 4 & 4 & 4 & 4 \\ \hline
    \textbf{2} & 10 & 10 & 10 & 10 & 10 & 10 & 10 & 8 & 4 & 7 & 5 & 1 & 1 & 9 \\ \hline
    \textbf{3} & 7 & 2 & 3 & 5 & 1 & 1 & 1 & 1 & 3 & 3 & 3 & 5 & 5 & 5 \\
    \hline
  \end{tabular}
\end{center}
\caption{Second plan $h$.}
   \label{tab:h}
\end{table}

\subsection{Optimization MAPF problem in a neighborhood of a reference plan}

After having defined these distances, we introduce a variant of the optimization MAPF problem (\ref{op}), the \textit{optimization MAPF problem constrained to a given plan}. We consider a sub-optimal solution $f_0$ of a MAPF instance, and we want to find another solution for the same problem which is not too far from $f_0$, and has better quality (i.e., shorter length). Let $\A$ and $\A^t$ be initial and final valid configurations on a digraph $G$. Let  $f_0 \in \L$ be such that $\A^t = \rho(\A,f_0) $ (i.e., $f_0$ is a feasible solution of the MAPF instance). Set $r \in {\mathbb{N}}$ and let $d$ be a distance between plans. The optimization MAPF problem, with \textit{Makespan}, constrained to $f_0$ is:
 
 \begin{equation} 
 	\label{opp}
 	\centering 
 	\begin{array}{ll}       	
 		
 		\min{|f|} & \\
 		
 		& \\
 		
 		\quad 	\text{s.t.} \;\; \A^t = \rho(\A,f) &  \\
 		
 		& \\
 		
 		\quad   \quad  \quad f \in \L, \;  d(f,f_0) \leq r. & \\
 		
 	\end{array}
 \end{equation}

 \section{Iterative local optimization for FSA}
 We present the basic idea of our approach in a slightly more general setting. Given a FSA $(\Sigma,X,x_0,\delta,X_M)$, consider problem
\begin{equation} 
	\label{op_gen}
	\centering 
	\begin{array}{ll}       	
		
	\min_{s \in \Sigma^*} {|s|} & \\
		
	\quad 	\text{s.t.} \;\; \delta(x_0,s)!\;\textrm{and}\;\delta(x_0,s) \in X_M. 
		
	\end{array}
\end{equation}

In Problem~(\ref{op_gen}), we want to find a string $s$ of minimum length such that FSA reaches a state in $X_M$.
Note that Problem~(\ref{op}) is a specific case of~(\ref{op_gen}),
with $\Sigma= \E$, $X=\C$, $x_0 =\A$, $X_M=\{\A^t\}$, $\delta=\rho$.
Let $d: \Sigma^* \times \Sigma^* \to \Real^+$ be a function such that

1) $d(s,s)=0$, $s \in \Sigma^*$,

2) $d(s\sigma,t)\geq d(s,t)$, $s,t \in \Sigma^*$, $\sigma \in \Sigma$,

3) $d(\epsilon,s)=0$, $s \in \Sigma^*$.

Roughly speaking, for $s,t \in \Sigma^*$, $d(s,t)$ represents the distance of $s$ to $t$. In 1) we require that the distance of $s$ to itself is $0$. In 2) we require $d(s,t)$ to be not smaller than $d(r,t)$, where $r$ is any prefix of $s$. In 3) we require the empty string $\epsilon$ to have $0$ distance to any string. Note that all distances defined in Section~\ref{sec:distances} satisfy these properties.
Let $s_0 \in \Sigma^*$ be a feasible solution of~(\ref{op_gen}) (that is,
$\delta(x_0,s_0) \in X_M$).
Let $\L$ be the subset of $\Sigma^*$ consisting of strings $s$ such that $\delta(x_0,s)!$.

Consider the following problem
 
  \begin{equation} 
 	\label{opp_gen}
 	\centering 
 	\begin{array}{ll}       	
 		
 		\min_{s \in \L} {|s|} & \\
 		
 		& \\
 		
 		\quad 	\text{s.t.} \;\;  \delta(x_0,s) \in X_M&  \\
 		
 		& \\
 		
 		\quad   d(s,s_0) \leq r. & \\
 		
 	\end{array}
 \end{equation}

Note that~(\ref{opp}) is a specific instance of~(\ref{opp_gen}) with $s_0=f_0$. The overall idea for solving~(\ref{op_gen}) consists of the following steps:
 \begin{itemize}
 \item Find a feasible solution $s_0$ of~(\ref{op_gen}). 
 \item Solve Problem~(\ref{opp_gen}). If the found solution $s^*$ is such that $|s^*|< |s|$, set $s_0=s^*$ and repeat this step.
\end{itemize}
 
We would like to solve Problem~(\ref{opp_gen}) by dynamic programming.
One problem is that the distance constraint $d(s,s_0)\leq r$ depends on string $s$, and it is not a function of the state only (that is, of $\delta(x_0,s)$).
To overcome this problem, we define an equivalence class $\sim$ on $\L$ such that, for any $s,t \in \L$ such that $s \sim t$, and any $\sigma \in \Sigma$
\begin{equation}
  \label{eqn_eq_class_prop}
  \begin{split}
     &|s|=|t|,\\
     &\delta(x_0,s)=\delta(x_0,t),\\
     &d(s \sigma,s_0) \leq r \Leftrightarrow  d(t \sigma,s_0) \leq r.
     \end{split}
   \end{equation}

 In other words, if $s \sim t$, then $s$ and $t$ have the same length and determine the transition of the FSA from the initial state $x_0$ to the same final state. Moreover, if $s\sigma$ satisfies the distance requirement, also $t \sigma$ satisfies it. 
 
 Then, we define another FSE $(\hat X,S,[\epsilon],\hat \delta,X_M/ \sim)$.
 The state space $\hat X$ is defined recursively as follows:

 1) $[\epsilon] \in \hat X$

 2) if $[s] \in \hat X$, $\delta(s, \sigma,s_0)!$, and $d(s \sigma,s_0)\leq r$ then
 $[s \sigma] \in \hat X$.

 Note that condition 2) is well-posed. Indeed, from our assumptions on $\sim$, condition $d(s \sigma,s_0) \leq r$ does not change if we substitute $s$ with any other member of equivalence class $[s]$.

We define transition function  $\hat \delta: \hat X \times S \to \hat X$ by setting

  1) $\hat \delta([s],\sigma)! \Leftrightarrow d(s \sigma,s_0) \leq r$,

2)  $\hat \delta([s],\sigma)=[s \sigma]$.

Then, we convert Problem~(\ref{opp_gen}) into the following
  \begin{equation} 
 	\label{opp_gen_red}
 	\centering 
 	\begin{array}{ll}       	
 		
 		\min_{s \in \L}{|s|} & \\
 		
 		& \\
 		
 		\quad 	\text{s.t.} \;\; \hat \delta([\epsilon],s) \in X_M / \sim.  \\
 		
 	\end{array}
 \end{equation}
 Note that in Problem~(\ref{opp_gen_red}), the state space is $\L / \sim$, the set of equivalence classes of $\L$ with respect to $\sim$. Problem~(\ref{opp_gen_red}) is a simple reachability problem for FSA, and can be solved by dynamic programming. Indeed, we removed from the new state space $\hat X$ all the states that correspond to strings that violate the distance constraint. Moreover, $\hat \delta$ allows only those transitions that lead to states inside $\hat X$. The following statement follows by construction.

 \begin{prop}
   String $s$ is a solution of~(\ref{opp_gen}) if and only if it is a solution of~(\ref{opp_gen_red}).
    \end{prop}


\subsection{Equivalence classes associated to the previously defined distances}
\label{eqrel}
We present the equivalence relations on $\L$ associated to the previously defined distances. These equivalence classes satisfy properties~(\ref{eqn_eq_class_prop}).
For the $\infty$-distance~(\ref{eq:inf_dist}), the max-min distance~(\ref{eq:max_min}), and   max-agents distance~(\ref{eq:dist_max_ag}), $f \sim g$ 
if and only if
	\begin{enumerate}
			\item $|f| = |g|$;
			
			\item $ \rho(\A,f)=\rho(\A,g)$. 
			
\end{enumerate}

We define an injective function $\alpha: \L / \sim\, \rightarrow {\mathbb{N}} \times \C$ as follows
\begin{equation}\label{a1}
\alpha([f])=(|f|,\rho(\A,f)).
\end{equation}
This function is well-defined because, if $f_1 \sim f_2$, then $|f_1|=|f_2|$ and $\rho(\A,f_1)=\rho(\A,f_2)$. Moreover, $\alpha$ is injective because, if $f_1$ and $f_2$ are such that $\alpha([f_1]) = \alpha([f_2])$, then $|f_1|=|f_2|$ and $\rho(\A,f_1)=\rho(\A,f_2)$, therefore, $[f_1] = [f_2]$.     
We will use function $\alpha$ to represent equivalence classes $\L / \sim$ in the dynamic programming solution algorithm.

For the $1$-distance~(\ref{eq:one_dist}) and the sum-min distance~(\ref{eq:sum_min}), $f \sim g$ if and only if

  \begin{enumerate}
		\item $|f| = |g|$;
			
		\item $ \rho(\A,f)=\rho(\A,g)$; 
			
		\item $d(f,f_0)= d(g,f_0)$;
	
                \end{enumerate}
                where $d=d_1$ for the $1$-distance and $d=d_1^*$ for the sum-min distance.
Again, we define an injective function $\alpha$, representing equivalence classes. Namely, $\alpha: \L / \sim\ \rightarrow {\mathbb{N}} \times \C \times {\mathbb{N}} $ is such that:
\begin{equation}\label{a2}
	\alpha([f])=(|f|,\rho(\A,f),d(f,f_0)).
      \end{equation}
We can show that this function is well-defined and injective with the same procedure used for~(\ref{a1}).

For the u-agents distance,~(\ref{eq:dist_ag}), $f \sim g$ if and only if

  \begin{enumerate}
		\item $|f| = |g|$;
			
                \item $ \rho(\A,f)=\rho(\A,g)$; 
		
        \item $\Q(f,f_0)= \Q(g,f_0)$;    

                \end{enumerate}
                where $\Q(f,f_0)$ is the subset of agents defined in (\ref{eq:Q_fg}).

We represent the state with $\alpha: \L / \sim\ \rightarrow {\mathbb{N}} \times \C \times {\mathbb{P}(P)} $, defined as follows: 
\begin{equation}\label{a3}
	\alpha([f])=(|f|,\rho(\A,f),\Q(f,f_0) ).
      \end{equation}
Again, we can show that this function is well-defined and injective, following the same procedure used for~(\ref{a1}).


\subsection{Neighborhoods}
\label{sec:neig}
Given the distances defined in Section \ref{sec:distances}, and the definition of the equivalence classes in Section \ref{eqrel}, we can define the neighborhood of a configuration and of an equivalence class. Moreover, we can upper estimate the cardinality of such neighborhoods.
Such an estimate is needed to evaluate the time needed to explore the neighborhoods, an operation that is central to the approach proposed in this paper.\\
We define a neighborhood of a configuration as a ball centered in $\A \in \C$ of radius $r$
\begin{equation}
\label{neig:conf}
\B_r(\A) := \{\A^* \in \C: \; d(\A^*,\A)\leq r\},
\end{equation}
where $d(\A^*, \A)$ can be either $\bar{d}(\A^*, \A)$ or $\tilde{d}(\A^*, \A)$.
Given a distance $d$ between plans and $r \in \mathbb{N}$, we define the neighborhood of an equivalence class $[f]\in \L$ as:
\begin{equation}
\label{neig:eq_class}
\B_r([f]) := \{[g], g \in \L: \; |g| \leq |f|, \,d(g,f) \leq r\}.
\end{equation}

\subsubsection{Paths Neighborhoods}

Let us consider the distances defined in Section~\ref{subsec:pathsdistances} and compute the upper bound for the cardinality of both neighborhoods defined in (\ref{neig:conf}) and (\ref{neig:eq_class}).
First, we define a ball $\B_r(v)$ centered at $v \in V$ of radius $r \in {\mathbb{N}}$:
\[\B_r(v) := \{u \in V: \; \hat d(u,v)\leq r\}.\]
Note that, for simplicity, we use the same notation ($\B_r$) for balls of vertices, configurations, and plans. The meaning of set $\B_r$ depends on the argument (vertex, configuration, or plan).
We denote by $\bar{\B}_r(v)$ the border of the ball, obtained by replacing $\leq$ with $=$ in the definition.
Let $\phi =outdeg(G)$ be the maximum out-degree of digraph $G=(V,E)$.
The following proposition provides an upper bound on the cardinality of $\bar{\B}_r(v)$.
\begin{prop}
\label{prop:dv}
It holds that
\begin{equation} \label{eq2}
		|\bar{\B}_r(v)| \leq  \phi^r.
\end{equation}
\end{prop}	
\begin{proof} 
Let  $n_h$ be the number of nodes at distance $h$ from $v$.  Note that $n_1 \leq \phi$, and $\forall h \geq 2$ $n_h \leq n_{h-1} (\phi -1)$. By induction, $n_h\leq \phi (\phi -1)^{h-1}$ $\forall h \geq 1$. Therefore, an upper-bound for the number of nodes on the border of the ball is 
\[|
  \bar{\B}_r(v)| \leq \phi (\phi -1)^{r-1} \leq \phi^r.
\]
\end{proof}
Next, we take into account the ball defined in (\ref{neig:conf}), denoting with $\bar{\B}_r(\A)$ its border.
An upper bound for the cardinality of the ball is given by the following proposition.
\begin{prop}
It holds that
\begin{equation}\label{eq3}
		|\B_r(\A)| \leq \; \binom{k+r}{r} \; \phi^r.
	\end{equation}
	where $k=|P|$ and $\phi=outdeg(G)$.
\end{prop}
\begin{proof} 
First of all, we find an upper bound for the number of configurations at distance $h$ from $\A$, i.e., an upper bound for the cardinality of the border of the ball centered in $\A$ of radius $h$:

\[ \bar{\B}_h(\A) = \left\{\A^* \in \C: \;\sum_{i=1}^{|P|} \hat d(\A(p_i),\A^*(p_i)) = h\right\}. \]

Let $(h_1, \cdots,h_{|P|})$ be a $|P|$-decomposition of $h$ (i.e., $\sum_{i=1}^{|P|} h_i = h $ ). Taking into account Proposition \ref{prop:dv}, an upper bound for the number of configurations for which $ \hat d(\A(p_i),\A^*(p_i)) = h_i$, $i=1,\ldots,|P|$, is
\[ \prod_{i=1}^{|P|}|\bar{\B}_{h_i}(\A(p_i))| \leq \prod_{i=1}^{|P|} \phi^{h_i}  = \phi^h. \]
By standard combinatorial arguments, the number of $|P|$-decompositions of $h$ is 
\[   \frac{(h + (|P|-1))!}{h! (|P|-1)!} .\]

Therefore, the cardinality of the border of the ball of radius $h$ can be bounded from above by:
\[|\bar{\B}_{h}(\A)| = \sum_{(h_1,\ldots,h_{|P|})\ :\ \sum_{i=1}^{|P|} h_i = h} \prod_{i=1}^{|P|}|\bar{\B}_{h_i}(\A(p_i))| \leq\]
\[\leq  \frac{(h + (|P|-1))!}{h! (|P|-1)!} \phi^h,\]
and the total number of configurations in $\B_{r}(\A)$ can be overestimated as follows: 
\[ |\B_{r}(\A)| = 1 +  \sum_{h=1}^{r}|\bar{\B}_{h}(\A)| \leq 1 +\sum_{h=1}^{r}  \frac{(h +|P|-1)!}{h! (|P|-1)!}  \phi^h =\]
\[ = \sum_{h=0}^{r}  \frac{(h +|P|-1)!}{h! (|P|-1)!}  \phi^h \leq \binom{|P|+r}{r}  \; \phi^r.\]
\end{proof}
From these bounds, we can compute an upper bound of the cardinality of $\B_r([f])$ for $f\in \L$ defined in (\ref{neig:eq_class}). Here, we consider the max-min distance (\ref{eq:max_min}). Since the max-min distance is the smallest among the path distances, the upper bound stated in the proposition also applies to the neighborhood constructed using the other path distances and, in particular, the sum-min distance (\ref{eq:sum_min}).
\begin{prop}
\label{prop:nr2f}
It holds that:
	\begin{equation}\label{eq5}
		|\B_r(f)| \leq   \,|f|^2 \; \binom{k+r}{r}  \; \phi^r \; (r+1),
	\end{equation} 
where $k = |P|$ and $\phi=outdeg(G)$. Moreover, $\exists \, C=C(r)\in (0,1]$ such that	
	\[|\B_r(f) | \leq |f|^2 \; C \; (k+r)^r \; \phi^r \; (r+1). \]	
\end{prop}
\begin{proof}
Let  $ f\in  \L$ and $\hat{f}\in  \B_r(f)$.
 Let $I = \{1, \cdots, |\hat{f}|\}  $, $J = \{1, \cdots, |f|\}$, and $R=\{0, \cdots, r\}$. Then:
\[\alpha(\B_r(f)) \subset  \left(I \times  \bigcup_{j \in J} \B_r(\psi_{f}(j)) \times R \right).\]
		        Indeed, it holds that $|\hat{f}| \leq |f|$ and, moreover, $d^*(\hat{f},f)\leq r$ implies that 
$$
\max_{1 \leq i \leq |\hat f|} \min_{1 \leq j \leq |\hat f|} \bar{d} (\psi_{\hat f}(i), \psi_{f}(j)) \leq r,
$$ 
and, in particular, for $i=|\hat{f}|$ there exists $j\in J$ such that
\[\bar{d} (\rho(\A, \hat{f}), \psi_{f}(j)) \leq r, \]
which means that $\rho(\A, \hat{f}) \in \B_r(\psi_{f}(j))$.  Therefore, recalling that $\alpha$ is injective,
\[|\B_r(f)| \leq \big|I \times  \bigcup_{j \in J} \B_r(\psi_{f}(j)) \times R \big| .\]
Then, also in view of  (\ref{eq3}), we have that:
\[ |\B_r(f) | \leq  |f| \left( \sum_{j=1}^{|f|}|\B_{r}(\psi_{f}(j))|  \right) \; (r+1) \leq \]
\[ \,\leq |f|^2 \; \binom{k+r}{r}  \; \phi^r \; (r+1).\]	
The last statement of the proposition follows from
\[ \binom{k+r}{r} \leq \frac{(k+r)^r}{r!},\]
and setting $C:=\frac{1}{r!}$.
\end{proof}
Note that given the last statement of the above proposition, we have that the neighborhood of $f$ of radius $r$ has a polynomial cardinality with respect to the number of nodes for each fixed integer value $r$.

\subsubsection{Agents Neighborhoods}

We consider the agent distances defined in Section~\ref{sec:a_d}, and compute upper bounds for the cardinality of the neighborhoods defined in (\ref{neig:conf}) and (\ref{neig:eq_class}).

First, we take into account the neighborhood of a configuration defined in (\ref{neig:conf}) with the agent distance, defined in (\ref{eq:dist_ag_conf}). The following result gives an upper bound on the cardinality of the corresponding ball.
\begin{prop}
\label{card_ag_1}
It holds that
\begin{equation}\label{eq3ag}
		|\B_r(\A)| \leq \binom{k}{r}\frac{(n-k+r)!}{(n-k)!},
\end{equation}
where $k = |P|$ is the number of agents and $n = |V|$ the number of nodes. 
\end{prop}
\begin{proof}
From the definitions of the ball in (\ref{neig:conf}) and of the distance in (\ref{eq:dist_ag_conf}), it follows that each configuration $\A^*$ in $\B_r(\A)$ has at most $r$ agents occupying different positions compared to configuration $\A$. We denote with $P_r \subset P$ the subset of $r$ agents that change their positions. The number of different choices of $r$ agents corresponds to the number of $r$-combinations of a set of cardinality $k$. It is the binomial coefficient $\binom{k}{r}$.
Since the remaining agents $P \setminus P_r$ are $k-r$, the available nodes that can be occupied by the agents in a given subset $P_r$ are
$n-k+r$.
So, for a fixed subset $P_r$, the new positions can be chosen in $\frac{(n-k+r)!}{(n-k)!}$ different ways.   
Therefore, the total number of different configurations $\A^*$  in $\B_r(\A)$ is
\[|\B_r(\A)| \leq \binom{k}{r}\frac{(n-k+r)!}{(n-k)!}.\]
\end{proof}
Then, we can compute an upper bound for the cardinality of $\B_r(f)$, with $f \in \L$, for the max-agents distance defined in (\ref{eq:dist_max_ag}).
\begin{prop}
	\label{neig3ag}
If we consider the max-agents distance~(\ref{eq:dist_max_ag}), it holds that:
	\begin{equation}\label{eq4ag_star}
		|\B_r(f)| \leq   \,|f| \,\binom{k}{r} \, \frac{(n-k+r)!}{(n-k)!}.
	\end{equation}
Moreover, we have that, for fixed $r$, the neighborhood has a polynomial cardinality with respect to the number of nodes. In particular, $\exists \, C=C(r)\in (0,1]$ such that	
	\[|\B_r(f) | \leq C \; |f| \; k^r \; (n-k+r)^r. \]	
\end{prop}
\begin{proof}
Let  $ f\in  \L$, $\hat{f}\in  \B_r(f)$, $I = \{1, \cdots, |\hat{f}|\}  $ and $J = \{1, \cdots, |f|\}  $. Let $${\cal P}_r=\{P_r\ :\ P_r\subset P,\ |P_r|=r\},$$ be the family of subset of $P$ of cardinality equal to $r$. Then:
 \begin{equation}
 \label{eq:alpha1_star}
\alpha(\B_r(f)) \subset  \left(I \times  \bigcup_{j \in J} \B_r(\psi_{f}(j)) \times {\cal P}_r \right).
\end{equation}
Indeed, it holds that $|\hat{f}| \leq |f|$ and $d_1^*(\hat{f},f)\leq r$ implies that, for $j=|\hat{f}|$,
 \[ \hat{d}(\rho^(\A,\hat{f}), \psi_{f}(|\hat{f}|)) = \hat{d}(\psi_{f}(|\hat{f}|), \psi_{f}(|\hat{f}|))\leq d_1^*(\hat{f},f) \leq r, \]
 which means that $\rho(\A,\hat{f}) \in  \B_r(\psi_{f}(|\hat{f}| ))$.
From (\ref{eq:alpha1_star}), note that:
 \[ \forall \hat{f} \in B_r(f) \; \; \alpha(\hat{f}) \in \{|\hat{f}|\} \times \B_r(\psi_{f}(|\hat{f}|)) \times {\cal P}_r.\] Therefore, $|\alpha(\hat{f})| \leq \big|\B_r(\psi_{f}(j))\times {\cal P}_r \big|$.
We can also state that, for a fixed $r$, the subset of agents whose position varies in $\psi_{f}(j)$ is a subset of the agents $i$ moved within path $f^i$. So the choice of the subset is already accounted for in the binomial coefficient inside the cardinality of $\B_r(\psi_{f}(j))$.
 
Recalling that $\alpha$ is injective, it follows that:
\[
|\B_r(f)| \leq |\alpha(\B_r(f))|\leq \big|\bigcup_{j \in J} \B_r(\psi_{f}(j)) \big| \leq \sum_{j=1}^{|f|}|\B_{r}(\psi_{f}(j))| \leq |f| \;\binom{k}{r}\frac{(n-k+r)!}{(n-k)!},	
\]	
where the last inequality follows from Proposition \ref{card_ag_1}.   
\newline\newline\newline
The last statement of the proposition follows from
\[ \binom{k}{r} \leq \frac{k^r}{r!},\ \ \ \frac{(n-k+r)!}{(n-k)!}=r!\binom{n-k+r}{r}\leq (n-k+r)^r,\]
after setting $C:=\frac{1}{r!}$.
\end{proof} 
We proved these results using the max-agents distance. However, from Proposition \ref{prop:dist} it follows that the upper bounds computed are also valid for neighborhoods defined with the u-agents distance (\ref{eq:dist_ag}). \\

\subsection{Iterative Neighborhood Search}
\label{sec:iter}
We propose an iterative approach for finding a sub-optimal solution of Problem~(\ref{opp_gen_red}). The returned solution is locally optimal with respect to the neighborhood of a reference solution. At each iteration, we solve an instance of Problem~(\ref{opp_gen_red}).
The algorithm takes as input a feasible solution $f_0$, that may be of poor quality. For instance, we can obtain $f_0$ from a rule-based algorithm, such as diSC \cite{ardizzoni2022}. We aim at improving $f_0$, obtaining a shorter solution.
To this end, we solve Problem~(\ref{opp_gen_red}) with a dynamic programming algorithm. For an assigned plan distance, we search in the neighborhood $\B_r(f_0)$ for plans shorter than $f_0$, through algorithm \emph{Dynprog}, which we will describe below.
If we cannot obtain a solution shorter than $f_0$ (that is, $f_0$ is locally optimal) we stop the algorithm. Otherwise, if we obtain an improved solution $f^*$, we redefine the reference solution as $f_0=f^*$ 
and repeat the whole procedure. We iterate until we cannot shorten the current solution any further.
This algorithm can be classified as a \textbf{Neighborhood Search} algorithm (see~\cite{pisinger2019,shaw1998}).

To define the neighborhood $B_r(f_0)$, we can use any distance function among those presented in Section~\ref{sec:distances}. In our numerical experiments, we used the sum-min distance and the u-agents distance defined in (\ref{eq:dist_ag}). Algorithm~\ref{alg1} presents the steps of the procedure just described.

\begin{algorithm}[H]
\caption{Neighborhood Search}
\label{alg1}

\begin{algorithmic}
\State Input: $f_0, \A, \A^t, r$
\State Output: $f^*$
\State $f^* \gets f_0$

\Do
\State $f_0 \gets f^*$
    \State $f^* \gets DynProg(f_0, B_r(f_0), \A, \A^t)$
\doWhile{$|f^*| < |f_0|$} 

\State \textbf{return} $f^*$
 
\end{algorithmic}
\end{algorithm}


\noindent

Also, we can alternate different types of neighborhoods.  This may allow us to find solutions of better quality, since we are actually exploring larger neighborhoods. For example, we propose Algorithm~\ref{alg_alt}, that first defines $\B_r(f_0)$ with the sum-min distance, and then defines it with the u-agents distance (\ref{eq:dist_ag}). In this case, the final solution is locally optimal with respect to both neighborhoods.

\begin{algorithm}[H]
\caption{Alternated Neighborhood Search}
\label{alg_alt}

\begin{algorithmic}
\State Input: $f_0, \A, \A^t, r$
\State Output: $f^*$
\State $f^* \gets f_0$

\Do
\State $f_0 \gets f^*$
    \State $f_1 \gets NeighSearch(f_0, \A, \A^t, \B_r(f_0))$, \textrm{using u-agents distance}
    \State $f^* \gets NeighSearch(f_1, \A, \A^t, \B_r(f_1))$, \textrm{using sum-min distance}
\doWhile{$|f^*| < |f_0|$} 

\State \textbf{return} $f^*$
 
\end{algorithmic}
\end{algorithm}
 
The input of this algorithm is made up of the initial plan $f_0$, the initial and final configurations $\A$, and $\A^t$ and the radius $r$. The output is the locally optimal solution $f^*$.

\subsection{Dynamic Programming Algorithm with Dominance}
\label{sec:DynProg_dom}
To search for the optimal solution to Problem (\ref{opp_gen_red}) we employ a \textit{Dynamic Programming} (DP) algorithm. In generic DP problems, we are given a state space $S$ where $A \subset S$ is the set of target states, a transition function $g: S \rightarrow P(S)$, where $P$ is the power set of $S$, and an objective function $c: S \rightarrow \mathbb{R}$. Starting from an initial state $s_0 \in S$, we iteratively expand states through a suitably defined transition function $g$, until we reach a state $s_t\in A$ with minimum objective function value.

\subsubsection{States and transition function for Path Distances}
When we employ the sum-min distance, the states represent the equivalence classes of relation $\sim$, defined in Section~\ref{eqrel} for this distance. Namely, we use the corresponding injection $\alpha: \L / \sim \rightarrow {\mathbb{N}} \times \C \times {\mathbb{N}}$ to associate to each equivalence class $[f]$ a triple $(\beta,\gamma,\sigma)=\alpha([f])$, where $\beta$ is the length of $f$, $\gamma$ the configuration obtained by applying $f$ to the initial state $\A$, and $\sigma$ is the distance of $f$ from reference plan $f_0$.
Namely, the state space is  
	\[\Ss := \alpha(\L / \sim) \subset {\mathbb{N}} \times \C \times {\mathbb{N}},\]
	where $\alpha$ is defined in (\ref{a2}).
Each state $s= (\beta, \gamma, \sigma) \in \Ss$ represents the equivalence class:
\[   \alpha^{-1}(s) = \{ f \in \L \, : \, \beta = |f|, \,\gamma = \rho(\A, f), \,\sigma = d^*_1(f,f_0) \}.\]  
The initial state is $s_0=\alpha(\epsilon)=(0, \A, 0)$.
We use a priority queue $Q$ to store the states that have not been visited yet. At the beginning, $Q=\{s_0\}$.
We define a partial ordering on $\Ss$ based on length. Namely, if $s_1=(\beta_1,\gamma_1,\sigma_1),s_2=(\beta_2,\gamma_2,\sigma_2)$, $s_1 < s_2$ if $\beta_1 < \beta_2$. We order the elements of $Q$ according to this ordering.

A state $s_1=(\beta_1,\gamma_1,\sigma_1)$ \emph{dominates} $s_2=(\beta_2,\gamma_2,\sigma_2)$ if
\[\beta_1 \leq \beta_2,\ \gamma_1=\gamma_2,\ \sigma_1 \leq \sigma_2.\]

In other words, $s_1$ dominates $s_2$ if the plans $f_1$, $f_2$, corresponding to $s_1$ and $s_2$, satisfy the following properties: Plan $f_1$ is not longer than $f_2$, $f_1$ and $f_2$ lead to the same final configuration, and the distance of $f_1$ from the reference solution $f_0$ is not larger than the one of $f_2$. If $s_1$ dominates $s_2$, we can discard $s_2$. In general, we remove from $Q$ all dominated states.

We also define the following {\em transition function}, which allows us to (possibly) add new states to the priority queue: 
\[\tilde{\rho}:\Ss \times \E \rightarrow \Ss  \]
\[ \tilde{\rho}((\beta,\gamma,\sigma), e) :=  (\beta + 1, \rho(\gamma, e), \sigma +   \min_{k \in {\mathbb{N}}} \bar{d}_\A(\rho(\gamma, e), \psi_{f_0}(k))). \]

Applying this function to state $s=(\beta, \gamma, \sigma)$:
\begin{itemize}
\item adds $1$ to length $\beta$,
\item updates the final configuration $\gamma$, applying function $\rho(\gamma, e)$ to $\gamma$, where $e$ is the chosen set of edges;
\item updates $\sigma$, adding the computed minimum distance between the updated final configuration $\rho(\gamma, e)$ and the reference plan $f_0$.

\end{itemize}  
We denote by
\[\Sigma:= \left\{\tilde{\rho}((\beta,\gamma,\sigma),e)\ :\  e \in \E\  \mbox{ and }\ \sigma +   \min_{k \in {\mathbb{N}}} \bar{d}_\A(\rho(\gamma, e), \psi_{f_0}(k))\leq r \right\}\subset \Ss,\]
the set of states generated through the transition function $\tilde{\rho}$, applied to current state $(\beta,\gamma,\sigma)$. 
Moreover, we denote with 
\[  \Gamma := \alpha(\B_r(f_0)) \subset \Ss,\]
the set of states that can be visited during a neighborhood search.

\subsubsection{States and transition function for Agent Distances}
\label{sub:ag_dist}

Using the u-agents distance introduced in Section~\ref{sec:a_d}, the states represent the equivalence classes of the corresponding relation $\sim$, presented in (\ref{a3}). 
We use the associated injection $\alpha: \L / \sim\, \rightarrow {\mathbb{N}} \times \C \times {\mathbb{N}}$ to assign, to each equivalence class $\hat f$, a triple $(\beta,\gamma,\Omega) = \alpha(\hat{f})$. Compared to the previous section, the third element of the triple is $\Omega$, which represents the subset of agents such that their path plans in a representative of $\hat f$ and in $f_0$ are different.

Namely, the state space is  
	\[\Ss := \alpha(\L / \sim) \subset {\mathbb{N}} \times \C \times {\mathbb{P}(P)},\]
	where $\alpha$ is defined in (\ref{a3}). Since $\alpha$ is injective, $\Ss $ and $\L / \sim$ are in one-to-one correspondence.
Each state $s= (\beta, \gamma, \Omega) \in \Ss$, represents the equivalence class:
\[   \alpha^{-1}(s) = \{ f \in \L \, : \, \beta = |f|, \,\gamma = \rho(\A, f),\Omega = \Q(f,f_0) \}.\]  
We set the initial state as $s_0=\alpha(\epsilon)=(0, \A, \emptyset)$ and use a priority queue $Q$ to store the states that will be explored. At the beginning, $Q=\{s_0\}$.
We order the elements of $Q$ following an ordering on $\Ss$ based on the length. Namely, if $s_1=(\beta_1,\gamma_1,\Omega_1),s_2=(\beta_2,\gamma_2,\Omega_2)$, $s_1 < s_2$ if $\beta_1 < \beta_2$.

A state $s_1=(\beta_1,\gamma_1,\Omega_1)$ \emph{dominates} $s_2=(\beta_2,\gamma_2,\Omega_2)$ if
\begin{itemize}
\item $\beta_1 = \beta_2$,
\item $\gamma_1=\gamma_2$,
\item $\Omega_1 \subseteq \Omega_2$.
\end{itemize}

In other words, all the representatives $f_1 \in \alpha^{-1}(s_1)$ and $f_2 \in\alpha^{-1}(s_2)$ satisfy the following properties:

\begin{itemize}

\item $|f_1|= |f_2|$:  $f_1$ is as long as than $f_2$;

\item $\rho(\A, f_1)= \rho(\A, f_2)$:  $f_1$ leads to the same configuration as $f_2$;

\item $\Q(f_1, f_0) \subseteq \Q(f_2, f_0)$: all the agents $p \in P$ such that $f_1^p \not = f_0^p$, are such that $f_2^p \not = f_0^p$.
\end{itemize}

We remove from $Q$ all dominated states. Again, we define a {\em transition function} $\hat{\rho}:\Ss \times \E \rightarrow \Ss $, which allows us to (possibly) add new states to the priority queue: 
\[ \hat{\rho}((\beta,\gamma,\Omega), e) :=  (\beta + 1, \rho(\gamma, e), \Omega \cup \Q(\rho(\gamma, e), \psi_{f_0}(\beta+1))). \]

Applying this function to state $s=(\beta, \gamma, \Omega)$:
\begin{itemize}
\item adds $1$ to length $\beta$,
\item updates configuration $\gamma$, applying $\rho(\gamma, e)$ to $\gamma$, where $e$ is the chosen set of edges (note that the choice for $e$ can be restricted to a subset of edges with cardinality not larger than $r$, and is, thus, polynomial for fixed $r$, due to the definition of the neighborhood);
\item updates $\Omega$, adding the agents that change their configuration from the reference plan in the step $\beta+1$ if they are not already in $\Omega$.

\end{itemize}  
As in the previous section, we denote by
$\Sigma$
the set of new states which can be generated through the transition function 
 and with $ \Gamma$ 
the set of states that can be visited during a neighborhood search.

\subsubsection{Algorithm}

Algorithm~\ref{dp_alg} describes the Dynamic Programming algorithm.
The priority queue $Q$ is a set of states, sorted according to the length $\beta$ of their representatives. The function ${\tt insert}(Q,x)$ inserts a state $x$ maintaining the partial order of $Q$, in the sense that, after the insertion of $x$, all the elements of the queue still respect the ordering 
	previously defined.
Function ${\tt remove}(Q,x)$ removes $x$ from $Q$.  The head of the queue, that is the state with minimal $\beta$, is denoted by $Q[0]$. The algorithm explores the state space starting from the initial state $s_0$. At each iteration, a state with minimum $\beta$ is extracted from the queue. If the state extracted is a target state, that is, its second component (the configuration) is $\A^t$, the algorithm stops, and we 
return a representative of the optimal solution of Problem  (\ref{opp_gen_red}) (for a given equivalence class $\hat{f}$, the function ${\tt repr}(\hat{f})$ returns a representative of the class).
Otherwise, the algorithm employs function ${\tt expand}(s,f_0,r)$, based on
the transition function previously defined, to generate new states. If a new state is not dominated, then it is added to the queue $Q$. Moreover, we remove from $Q$ all states dominated by one of the newly added states.\\
\begin{algorithm}[H]
\caption{Dynamic Programming with Dominance}
\label{dp_alg}
\begin{algorithmic}

\State Input: $f_0, r, \A, \A^t$
\State Output: $f$
\BState $s_0 \gets (0, \A, 0)$ or $(0, \A, \emptyset)$
\State ${\tt insert}(Q, s_0)$
\While{$Q \neq \emptyset$}:

	\State $s=(\beta,\gamma,\sigma) \gets Q[0]$
	\If{$\gamma=\A_t$}
		\State $f \gets {\tt repr}(\alpha^{-1}(s))$   
\State $Q \gets \emptyset$
 \Else
	\State $\Sigma \gets {\tt expand}(s, f_0, r)$
	\For{$s_k \in \Sigma$}
		\If{$s_k$ is not dominated in $Q$}
			\State ${\tt insert}(Q, s_k)$ 
			\For{$s_i \in Q$}
				\If{$s_k$ dominates $s_i$}
					\State ${\tt remove}(Q, s_i)$
				\EndIf	
			\EndFor
		\EndIf
	\EndFor
\EndIf
\EndWhile
\State
\State \textbf{return} $f$
\end{algorithmic}
\end{algorithm}

\subsection{Complexity results}
In this section we provide a complexity analysis of the proposed algorithms. 
\begin{theorem}
  \label{thm_complx}
 Algorithms \ref{alg1}, \ref{alg_alt} and \ref{dp_alg} have polynomial time complexity with respect to the number of nodes of the graph, for a fixed value $r$, and assuming that each call of the procedure {\tt expand} has unit cost.
\end{theorem}

\begin{proof}
First, we consider the case with the sum-min distance.  
At each iteration of Algorithm \ref{dp_alg} we remove one state $s$ from set $Q$. Since $Q\subseteq \Gamma$, an upper bound for the number of iterations is
$|\Gamma|$. If the state $s$ is not a final state, then it is expanded and generates a set $\Sigma$ of new states. For each newly generated state, we need to check whether it is dominated from or dominates states that are already in $Q$.
A rough implementation of such check operation requires $O(|Q|)$ operations. Since the check must be repeated for all states in $\Sigma$, the overall number of operations in a single iteration is $O(|Q||\Sigma|)$.
Since it also holds that $\Sigma\subseteq \Gamma$, recalling the bound on the number of iterations, we have that the overall number of operations of the algorithm is $O(|\Gamma|^3)$.
Reminding that $\alpha$ is injective and using the result of Proposition \ref{prop:nr2f},
\[ |\Gamma|= |\B_r(f_0)|\leq |f_0|^2 C (k+r)^r\phi^r \; (r+1), \]	
where $k=|P|$ and $\phi=outdeg(G)$. Therefore, the time complexity of Algorithm \ref{dp_alg} is
\[O(|f_0|^6 \; C^3 \; (k+r)^{3r} \; \phi^{3r} \; (r+1)^3).\]
Algorithm \ref{alg1} calls Algorithm \ref{dp_alg} at most $|f_0|$ times (after each call the length of the plan is decreased by at least one), and so it has time complexity \[O(|f_0|^7 \; C^3 \; (k+r)^{3r} \; \phi^{3r} \; (r+1)^3).\]\\
Polynomiality, for fixed $r$, follows from the observation that rule-based algorithms are able to return an initial plan $f_0$ with polynomial length.
\newline\newline\noindent
For the case where the agent distance defined in  (\ref{eq:dist_ag}) is used, a similar development, exploiting Proposition \ref{neig3ag}, leads to the complexity $O(|f_0|^3 \; C^3 \; k^{3r} \; (n-k+r)^{3r})$ for Algorithm \ref{dp_alg}, and $O(|f_0|^4 \; C^3 \; k^{3r} \; (n-k+r)^{3r})$ for Algorithm \ref{alg1}.\newline\newline\noindent
Finally, we observe that also Algorithm \ref{alg_alt} has a polynomial time complexity, as it calls Algorithm \ref{alg1} with the two different neighborhoods at most $|f_0|$ times.
\end{proof}
Then, polynomial time complexity of Algorithms \ref{alg1}, \ref{alg_alt} and \ref{dp_alg} is established if we are able to prove that the {\tt expand} procedure has polynomial time complexity. This will be shown in what follows.
\subsubsection{Procedure {\tt expand}}
As explained in the proof of Theorem \ref{thm_complx}, the {\tt expand} function returns a number of states $\Sigma \subset \Gamma$ that grows polynomially with the number of agents. The set of states reachable from a state $s_0$, without keeping into account the distance constraint is:
\[
\Sigma_0 = \left\{\tilde{\rho}(s_0,e)\ :\  e \in \E \right\},
\]
whose cardinality, in the worst case, grows exponentially with respect to the number of agents. For instance, if the out-degree of all the nodes in $\gamma$ is $\phi$ (the maximum out-degree of the graph), the number of configurations reachable from $\gamma$ could be as large as $\phi^{|P|}$ (recall that $\E=|\bar{E}|^{|P|}$). This means that we cannot compute $\Sigma$ by first computing $\Sigma_0$ and then discarding all its elements with
distance larger than $r$ from $f_0$, as the resulting time complexity would be exponential with respect to the number of agents. We can use different approaches, depending of the type of distance used, to overcome this problem.

\paragraph{U-Agents Distance}
In case the U-Agents Distance is employed, and the state is $s_0=(\beta, \gamma, \Omega)$, the {\tt expand} procedure is polynomial as the search for new configurations can be performed as follows:
\begin{itemize}
\item First, we choose the agents that will change their path with respect to the original path $f_0$. In the worst case, $\Omega$, that is the set of agents that have already changed their path, is empty, so we have to choose $r$ agents. The number of different ways we can choose the agents is $\binom{|P|}{r}$.  
\item Then, we search for the new positions of those agents. As the agents that change their positions are $r$, in the worst case the number of configurations reachable from $\gamma$ is $\phi^r$, where $\phi$ is the maximum out-degree of the graph.
\end{itemize}
With this implementation, an upper bound for the time complexity of the {\tt expand} procedure is:
\[
\binom{|P|}{r} \; \phi^r \leq \frac{|P|^r}{r!} \; \phi^r,
\]
and is thus polynomial with respect to the number of agents (for fixed $r$).

\paragraph{Sum-Min Distance}

If we employ the sum-min distance, we introduce Algorithm \ref{alg:expand} as a possibile implementation of the procedure {\tt expand}. 
The main idea is to update the position of one agent at a time, checking for each agent that the configurations we are building are inside the neighborhood. 
A configuration $\A$ of ordered agents in $P$ can be seen as a string $a_{|P|}=n_1n_2\ldots,n_{|P|}$ such that
$\A(q)=n_q$ for $q=1,\ldots,|P|$. Next, we can define a partial configuration with length $t\leq |P|$ as a string $a_t=n_1n_2\ldots n_t$. Finally, for a given plan $f_0$, we can define the distance
between the partial configuration $a_t$ and plan $f_0$ as follows:
$$
d(a_t,f_0)=\sum_{i=1}^t    \min_{k \in \{1,\ldots,|f_0|\}  } \hat d(n_i, (\psi_{f_0}(k))_i),
$$
where $(\psi_{f_0}(k))_i$ is the $i$-th symbol of the string $\psi_{f_0}(k)$.

\begin{algorithm}[H]
\caption{Expand for Path Distance}
\label{alg:expand}
\begin{algorithmic}[1]

\State Input: $f_0, r, s_0 = (\beta, \gamma, \sigma)$
\State Output: $\Sigma$
\State $L_1 \gets \{\epsilon\}$
\For{$p \in P$}
\State $N \gets adjacent\_nodes(\gamma(p))$
\State $L_0 \gets L_1$
\State $L_1 \gets \varnothing$
\For{$a_0 \in L_0$}
\For{$n \in N$}
		\State $a \gets a_0 n$		\label{l0}
		\If {$\sigma + d(a, f_0) \leq r \wedge  isvalid(a)$} \label{l1}
			\State $L_1 \gets L_1 \cup \{a\}$
		\EndIf
\EndFor
\EndFor
\EndFor
\vspace{7pt}
\State $\Sigma \gets \varnothing$
\For{$a \in L_1$}
\State $s \gets \left(\beta+1, a, \sigma + d(a,f_0)\right)$
\State $\Sigma \gets \Sigma \cup \{s\}$

\EndFor 

\State \textbf{return} $\Sigma$
\end{algorithmic}
\end{algorithm}

Algorithm \ref{alg:expand} takes in input the original plan $f_0$, the distance $r$ and the state $s_0=(\beta, \gamma, \sigma)$ and returns a list of states $\Sigma$. 
The algorithm first initializes a list $L_{1}$ of strings with the empty string $\epsilon$. 
Then, for each $p\in P$, we first generate the set $N$ of nodes adjacent to node $\gamma(p)$ (procedure $adjacent\_nodes$). Next, we copy $L_1$ into the list $L_0$ and we set $L_1$ equal to the empty collection.
Then, for each partial configuration $a_0\in L_0$ and each node $n\in N$, we generate a new string $a$ appending $n$ at the end of $a_0$. If the new partial configuration $a$ does not lead to a violation of the distance constraint,
i.e., the sum of $\sigma$ and $d(a,f_0)$ does not exceed $r$, and, moreover, the partial configuration $a$ is valid (procedure $isvalid$ checks that each node inside $a$ is not repeated, i.e., that each node is occupied by at most one agent), we add $a$ to $L_1$.
Once all agents have been processed and $L_1$ only contains full configurations, we generate a new state with first component $\beta+1$ for each member of $L_1$, and we add it to $\Sigma$, which is then returned by the procedure.

\begin{prop}\label{prop:expand}
Let $\mathbb{A}_P$ be the set of all the partial configurations 
generated at line \ref{l0} of Algorithm~ \ref{alg:expand}. 
Then,
\[|\mathbb{A}_P| \leq \sum_{k=0}^{|P|-1} \phi^{r+1}\binom{k+r}{r} \; |f_0|.\]
Moreover, $\exists C = C(r) \in (0, 1]$ such that:
\[
|\mathbb{A}_P| \leq C \; |P| \; \phi^r \; (|P|-1+r)^r.
\]
\end{prop}

\begin{proof}
For an ordered subset of agents $I=\{1,\ldots,i\}\subseteq P$, we denote with $\mathbb{A}_I$ the set of partial configurations of length in $\{1, \ldots, i\}$
generated up to iteration $i$ at line \ref{l0} of Algorithm \ref{alg:expand}.
 We can prove that:
\begin{equation}\label{eq:A_I}
|\mathbb{A}_I| \leq \sum_{k=0}^{i-1} \phi^{r+1}\binom{k+r}{r} \; |f_0|.
\end{equation}
This can be achieved by induction.
\begin{itemize}
\item \emph{Base Case}:
for $i=1$ we have that the number of partial configurations $|\mathbb{A}_{\{1\}}|$ is simply the number of the nodes adjacent to the position of agent 1, which is not larger than $\phi$, where $\phi$ is the maximum out-degree of the graph.
Then:
\[
i=1 \implies |\mathbb{A}_1|\leq\phi \leq  \phi^{r+1}\; |f_0|=         \sum_{k=0}^{1-1} \phi^{r+1}\binom{k+r}{r} \; |f_0|=
\phi^{r+1}\binom{r}{r} \; |f_0|.
\]
\item \emph{Induction Step}: Now we assume that the statement is true for $I$ and we prove it is also true for $I\cup \{i+1\}$.
By the inductive assumption it holds that:
\[|\mathbb{A}_I| \leq \sum_{k=0}^{i-1} \phi^{r+1}\binom{k+r}{r} \; |f_0|.\]
Then, we can compute the number of partial configurations explored for agent $i+1$. Looking at Algorithm \ref{alg:expand}, all partial configurations that are in the list $L_0$ when we add a node adjacent
to the $(i+1)$-th node of the current configuration $\gamma$, have a distance from $f_0$ lower than or equal to $r$.  Let $f^I_0 \in \E^*$ be the partial plan of agents $\{1,\ldots, i\}$, i.e., plan $f_0$ restricted to the first $i$ agents.
Recalling the definition of the neighborhood of a configuration in (\ref{neig:conf}), $L_0$ is a subset of the union of all the neighborhoods of the partial configurations for agents $\{1,\ldots, i\}$ inside $f^I_0$. We have that:
\[
|L_0| \leq  \left|\bigcup_{j=1,\ldots,|f^I_0|} \B_r(\psi_{f^I_0}(j))\right| \leq \phi^r \; \binom{i+r}{r} \; |f_0|.
\]

For each partial configuration inside $L_0$, we add the new possible positions of agent $i+1$ to create all the new partial configurations for agents $\{1,\ldots, i+1\}$. As the new possible positions of agent $i+1$ are at most $\phi$, the set $\mathbb{S}_{i+1}$ of new partial configurations is such that: 

\[
|\mathbb{S}_{i+1}| \leq \phi^{r+1} \; \binom{i+r}{r} \; |f_0|.
\]
Therefore, at the end of step $i+1$, the number of partial configurations explored is:
\[
|\mathbb{A}_{I\cup\{i+1\}}|=|\mathbb{A}_I|+|\mathbb{S}_{i+1}| \leq \sum_{k=0}^{i-1} \phi^{r+1}\binom{k+r}{r} \; |f_0| + \phi^{r+1} \;  \binom{i+r}{r} \;|f_0| = \sum_{k=0}^{i+1-1} \phi^{r+1}\binom{k+r}{r} \; |f_0|.
\]
\end{itemize}
The proposition follows directly from (\ref{eq:A_I}) when $I=P$.
\newline\newline\noindent
The last statement of the proposition follows from 
\[
\sum_{k=0}^{|P|-1} \phi^{r+1}\binom{k+r}{r} \; |f_0| \leq \sum_{k=0}^{|P|-1} \phi^{r+1}\frac{(k+r)^r}{r!} \; |f_0| \leq |P| \;  \phi^{r+1}\frac{(|P|-1+r)^r}{r!},
\]
and setting $C := \frac{1}{r!}$.

\end{proof}

\begin{prop}
Algorithm \ref{alg:expand} has a polynomial time complexity with respect to the number of agents for fixed $r$. 
\end{prop}

\begin{proof}
The time complexity of Algorithm \ref{alg:expand} follows from Proposition \ref{prop:expand} and the observation that the most expensive operation at the $i$-th iteration, i.e., the computation, performed at line \ref{l1}, of
the distance $d$ between each partial configuration with length $i$ and plan $f_0$, requires $i|f_0|$ operations. Indeed, for each agent $j\in\{1,\ldots,i\}$ we need to compute the minimum distance between 
the $j$-th node of the partial configuration and node $(\psi_f(k))_j$, for $k\in \{1,\ldots,|f_0|\}$.
\end{proof}

\section{Experimental results}
In this section, we test the local search algorithms presented in the paper. In particular, we performed experiments with Algorithm \ref{alg1}, both with path distances and agent distances, and with the alternating Algorithm \ref{alg_alt}.  

We coded Algorithm~\ref{alg1} and Algorithm~\ref{alg_alt} in \textit{C++}. We ran all tests on a \textit{11th Gen Intel(R) Core(TM) i7-1165G7 @ 2.80GHz} processor with a 16 GB RAM.

We generated random directed graphs with $|V|$ nodes ranging from 20 to 100 by 10, and $|E|=4|V|$ edges. The graphs are generated by creating $|E|$ random ordered pair of nodes and using them to build a directed graph. Only strongly connected graphs are selected. The number of agents $|P|$ ranges from 2 to 18, while initial and final configurations are randomly generated. Initial solutions (if they exist) are generated through the diSC algorithm described in \cite{ardizzoni2022}. Note that this algorithm is complete, i.e., it always returns a feasible solution in case one exists. However, the quality of such an initial solution is usually low (i.e., the length of the plan is usually long). For each set of experiments we show the reduction ratio obtained after the application of local search, compared to the initial solution, and the running time (in seconds). The reduction ratio is given by the ratio $\frac{|f^*|}{|f_0|}$, where $f_0$ and $f^*$ are, respectively, the initial and the final solution. The graphs reported show the results wrt to $|V|$ for $|P|=\{5,10,15\}$ and wrt to $|P|$ for $|V|=\{20,50,100\}$.

\subsection{Path Distance}
In the first set of experiments we teseted Algorithm \ref{alg1} with the sum-min distance. As the time needed to explore the neighborhood increases exponentially with the radius $r$, we set $r=1$ to reduce computational times.
\begin{figure}[H]
\begin{multicols}{3}
\centering
\includegraphics[width=0.33\textwidth]{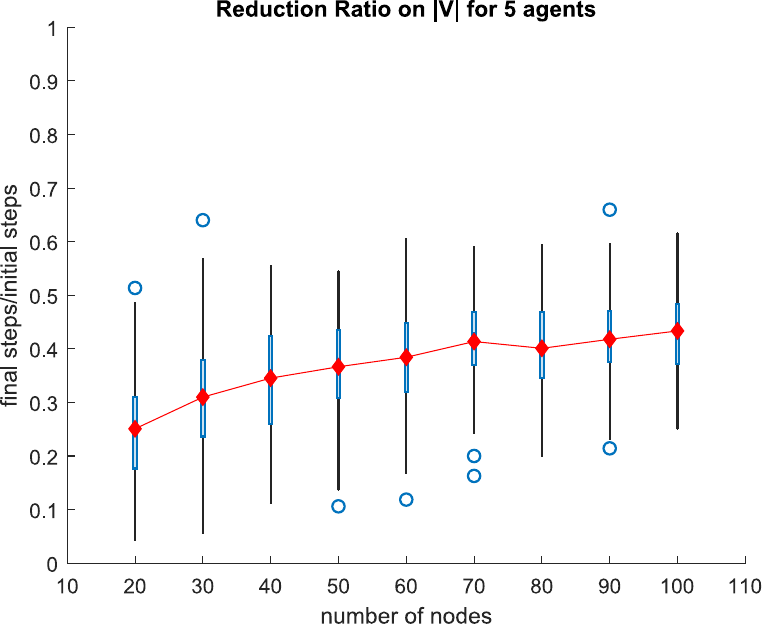}\par
\includegraphics[width=0.33\textwidth]{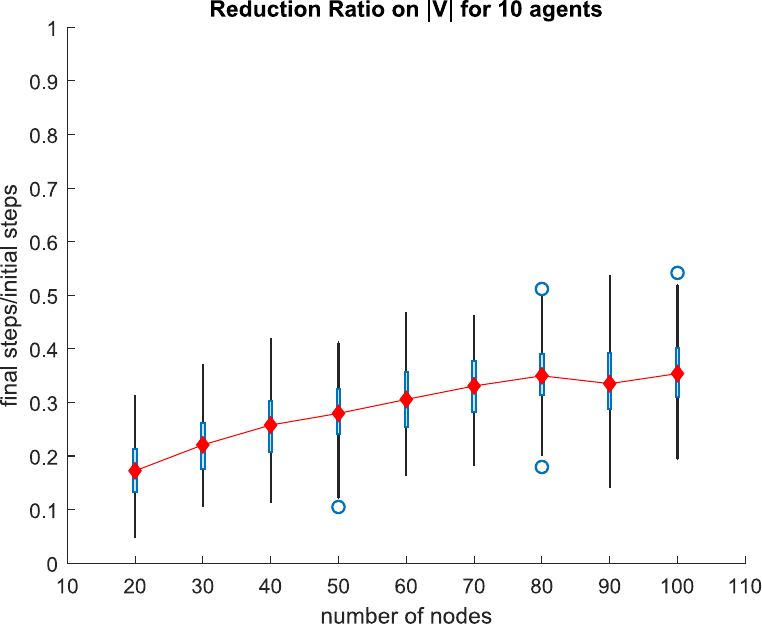}\par
\includegraphics[width=0.33\textwidth]{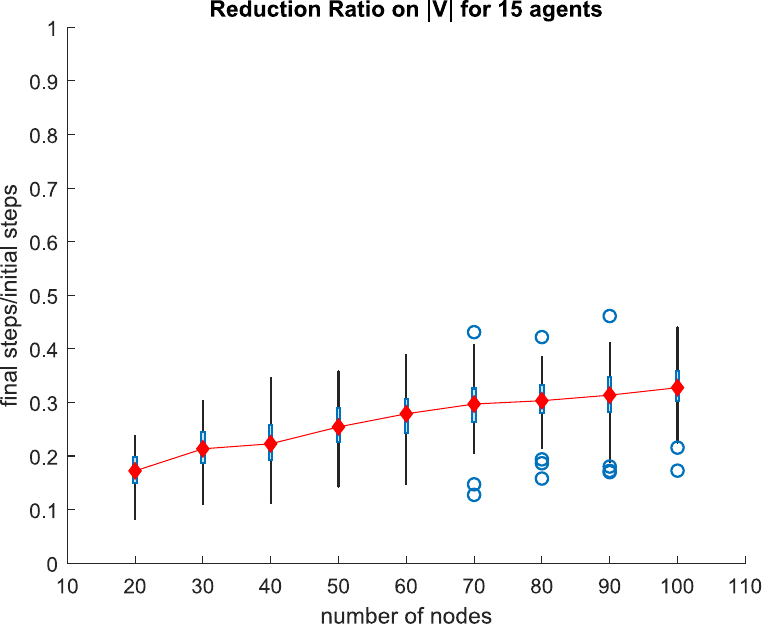}\par
\end{multicols}
\caption{Average Reduction Ratio per $|V|$.}
\label{fig:nodes_rr_ls}
\end{figure}
\begin{figure}[H]
\begin{multicols}{3}
\centering
\includegraphics[width=0.33\textwidth]{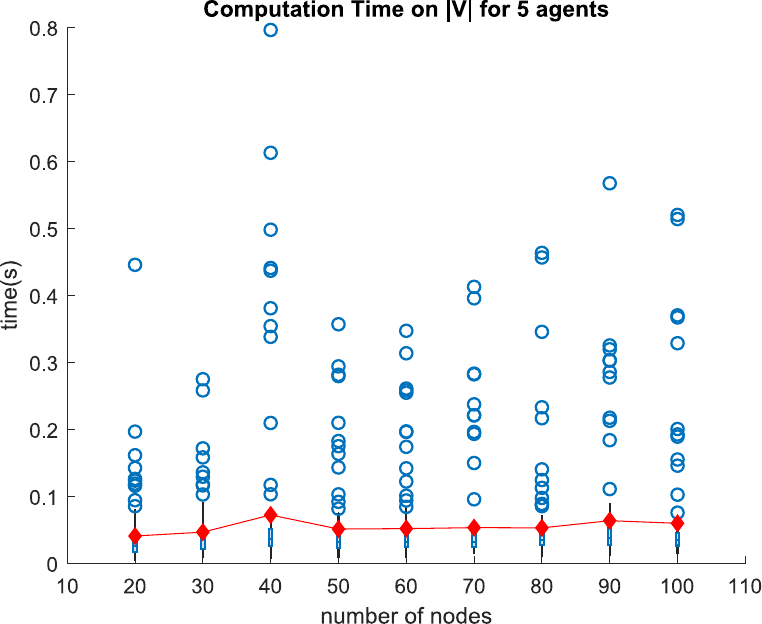}\par
\includegraphics[width=0.33\textwidth]{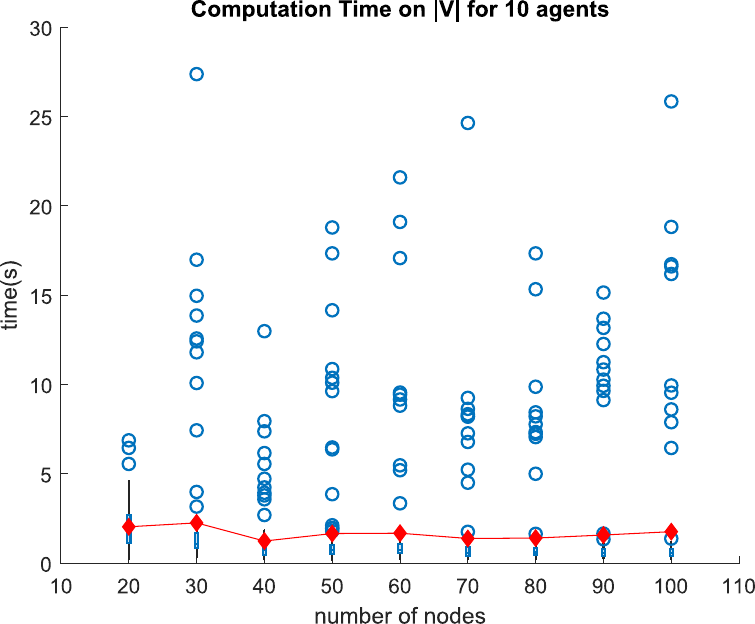}\par
\includegraphics[width=0.33\textwidth]{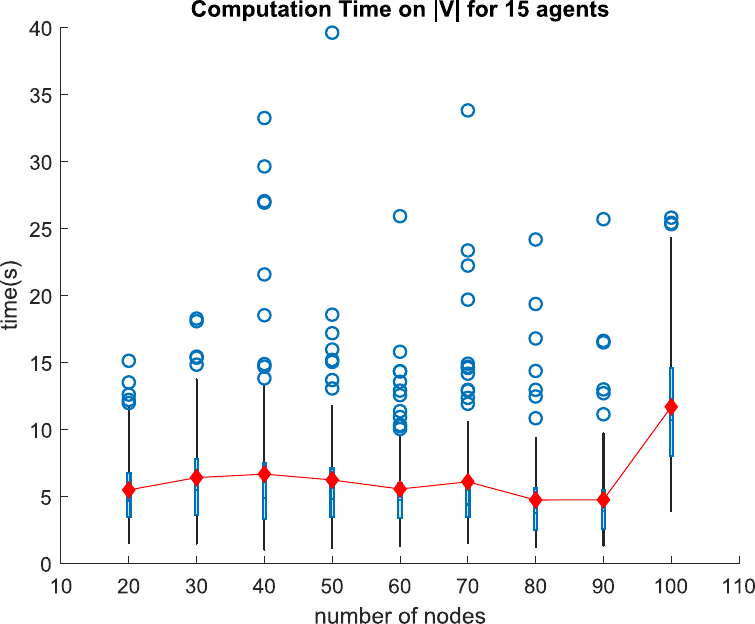}\par
\end{multicols}
\caption{Average Running Time per $|V|$.}
\label{fig:nodes_t_ls}
\end{figure}
 
Figure~\ref{fig:nodes_rr_ls} and Figure~\ref{fig:nodes_t_ls} show the reduction ratio (that is, the ratio between the lenghts of the final solution and the initial one) and the running time of Algorithm~\ref{alg1} wrt to the number of nodes $|V|$, for three different numbers $|P|$ of agents  (namely, $|P|=5,10,15$). The red line represents the average. In Figure~\ref{fig:nodes_rr_ls} the reduction ratio slightly increases as the number of nodes grows, for all the three different numbers of agents $|P|$. Instead, the computation time remains nearly constant.

\begin{figure}[H]
\begin{multicols}{3}
\centering
\includegraphics[width=0.33\textwidth]{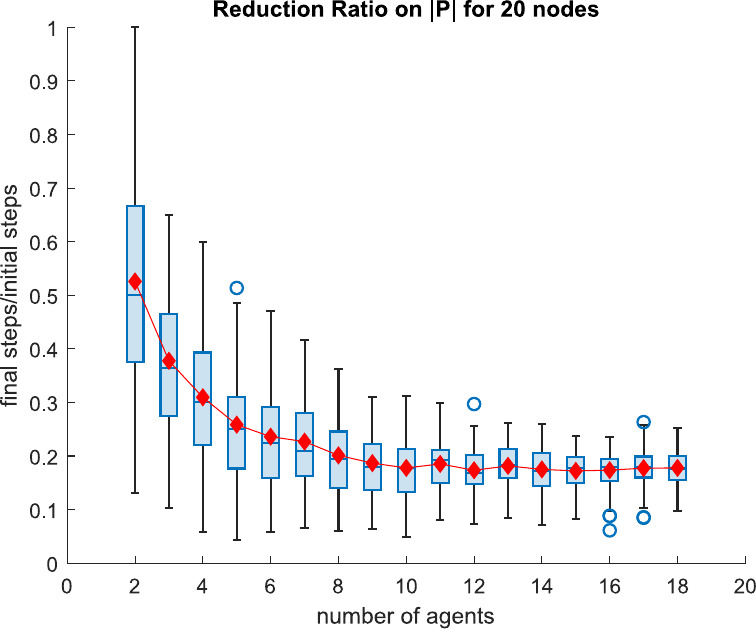}\par
\includegraphics[width=0.33\textwidth]{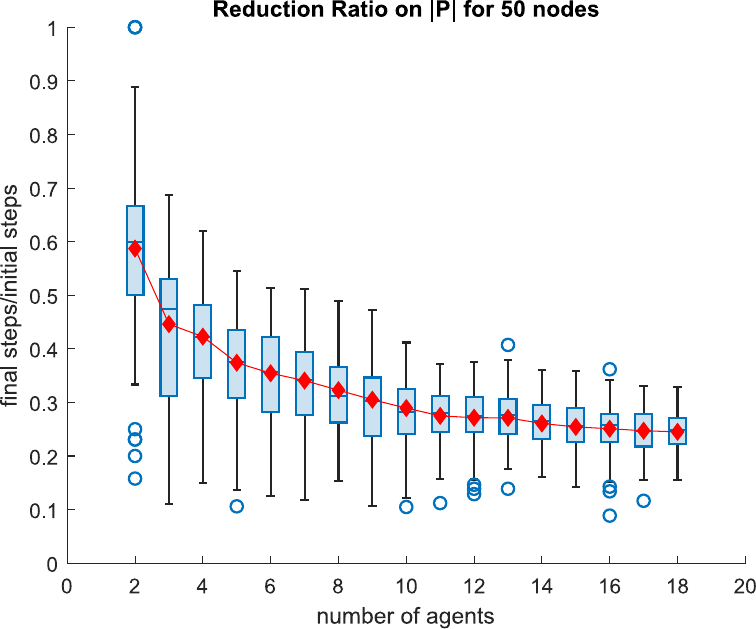}\par
\includegraphics[width=0.33\textwidth]{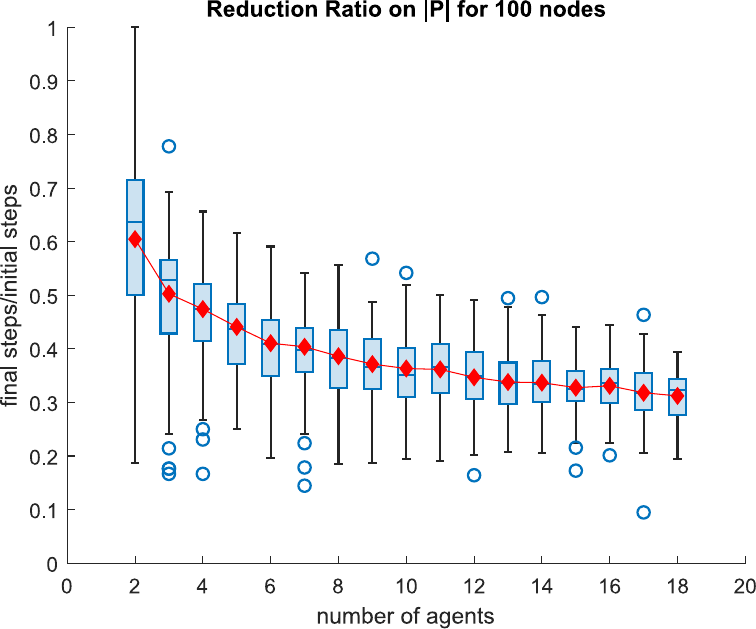}\par
\end{multicols}
\caption{Average Reduction Ratio per $|P|$.}
\label{fig:agents_rr_ls}
\end{figure}
\begin{figure}[H]
\begin{multicols}{3}
\centering
\includegraphics[width=0.33\textwidth]{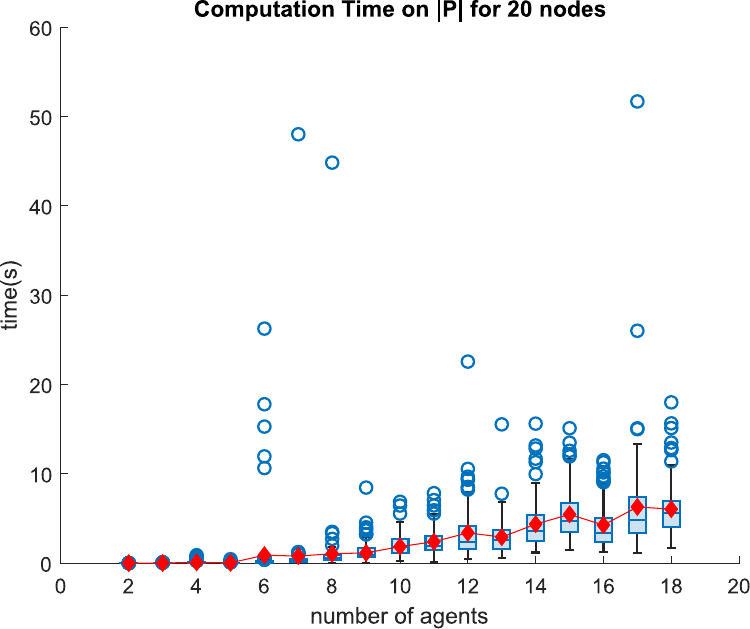}\par
\includegraphics[width=0.33\textwidth]{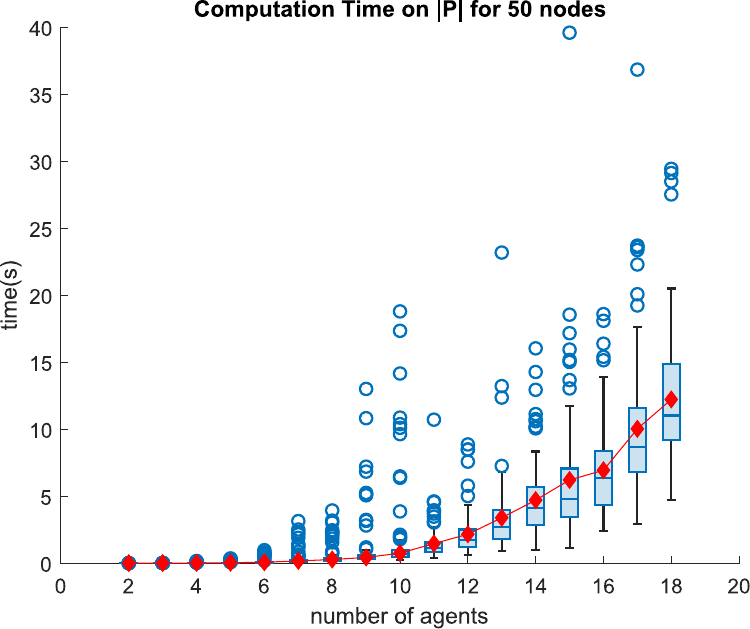}\par
\includegraphics[width=0.33\textwidth]{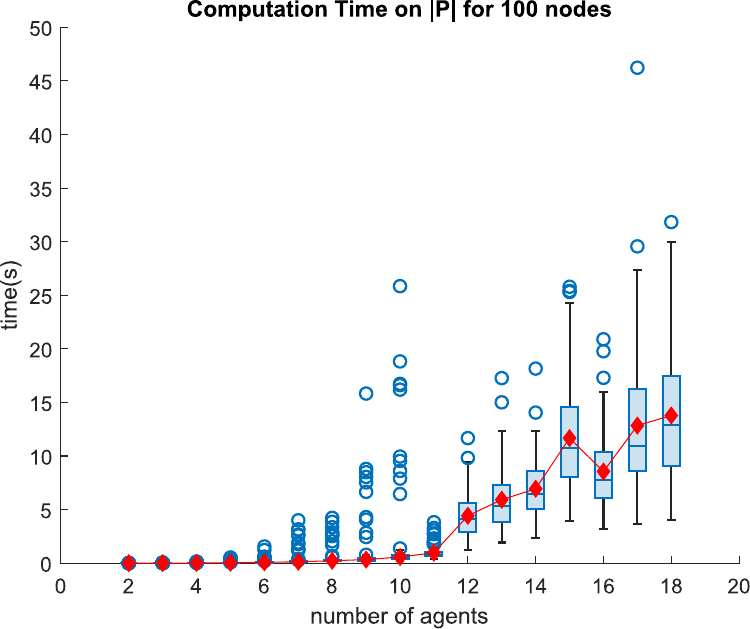}\par
\end{multicols}
\caption{Average Running Time per $|P|$.}
\label{fig:agents_t_ls}
\end{figure}
In Figure~\ref{fig:agents_rr_ls} and Figure~\ref{fig:agents_t_ls}, we report the reduction ratio and the running time (in seconds), respectively, wrt to $|P|$, for three values of $|V|$  (namely, $|V|=20,50,100$). The average is highlighted in red. We note that the reduction ratio decreases with the number of agents, while the computation time increases. The trends in the plots, wrt both the number of nodes and the number of agents, suggest that a larger improvement is more likely for more complex instances, i.e., those with a limited difference between the number of nodes and the number of agents, which makes the graph more crowded.

\subsection{Agent Distance}
In the second set of experiments, we tested Algorithm~\ref{alg1} with u-agents distance. We set the radius $r=1$. That is, at each step, the local optimization can change the path of at most one agent, with respect to the previous soution.
\begin{figure}[H]
\begin{multicols}{3}
\centering
\includegraphics[width=0.33\textwidth]{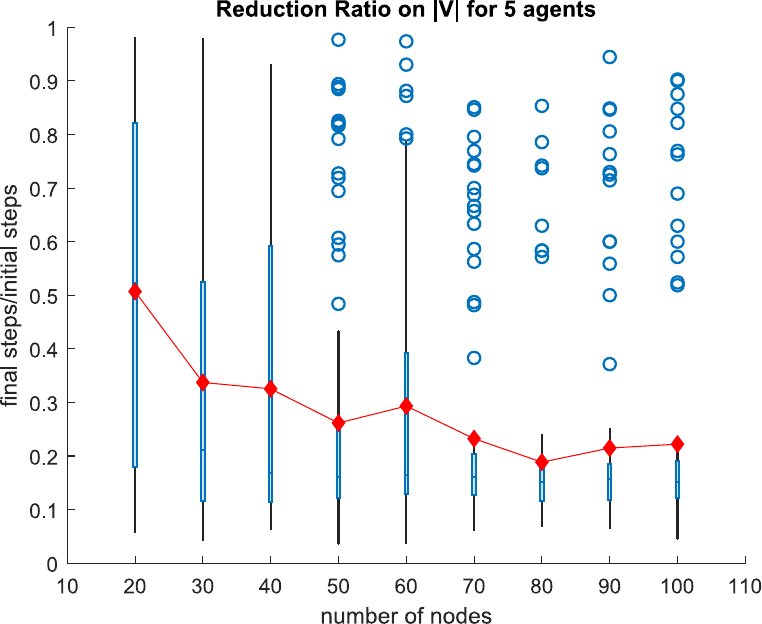}\par
\includegraphics[width=0.33\textwidth]{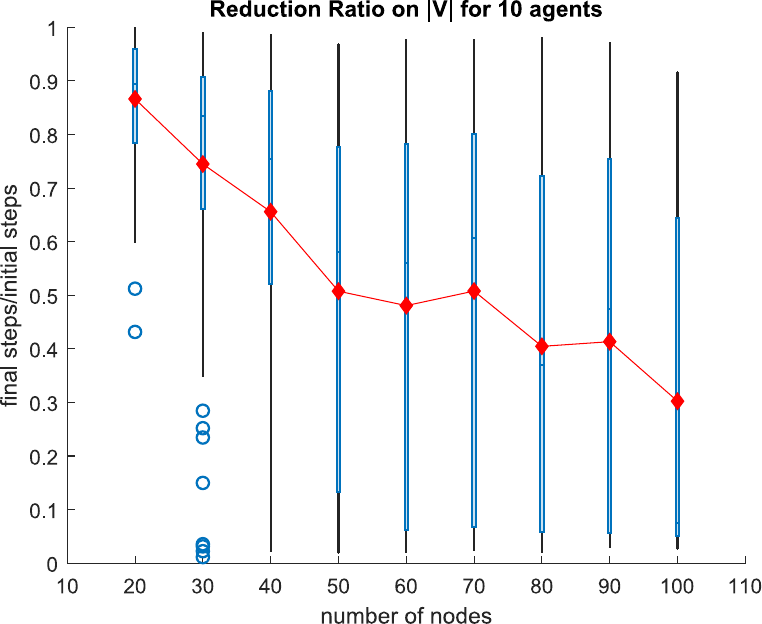}\par
\includegraphics[width=0.33\textwidth]{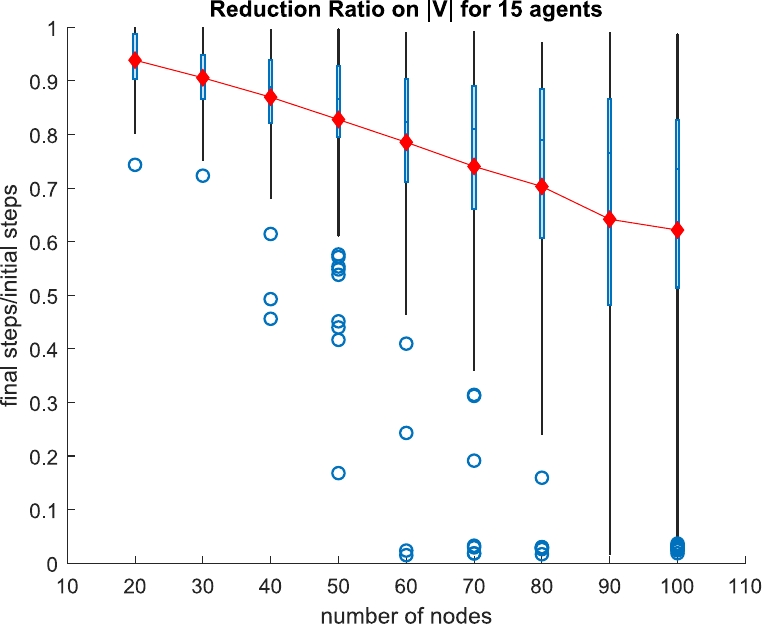}\par
\end{multicols}
\caption{Average Reduction Ratio per $|V|$.}
\label{fig:nodes_rr_tw}
\end{figure}
\begin{figure}[H]
\begin{multicols}{3}
\centering
\includegraphics[width=0.33\textwidth]{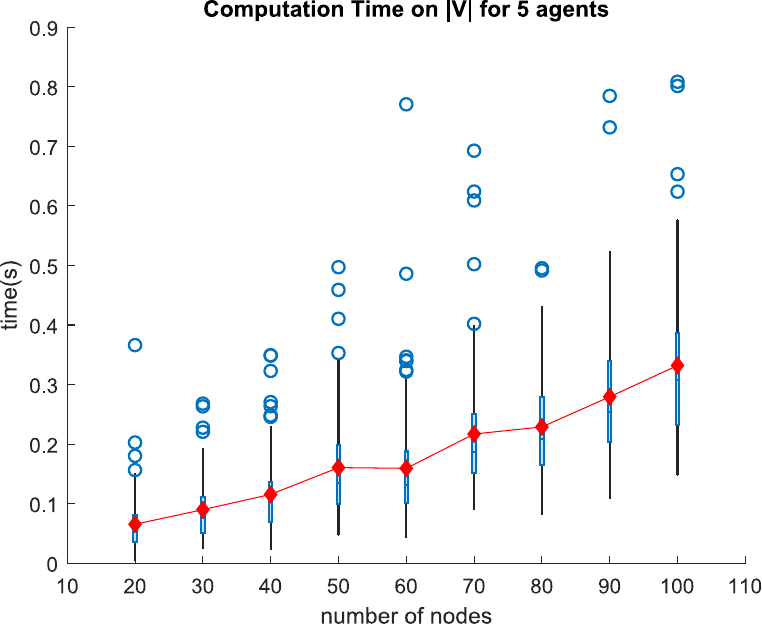}\par
\includegraphics[width=0.33\textwidth]{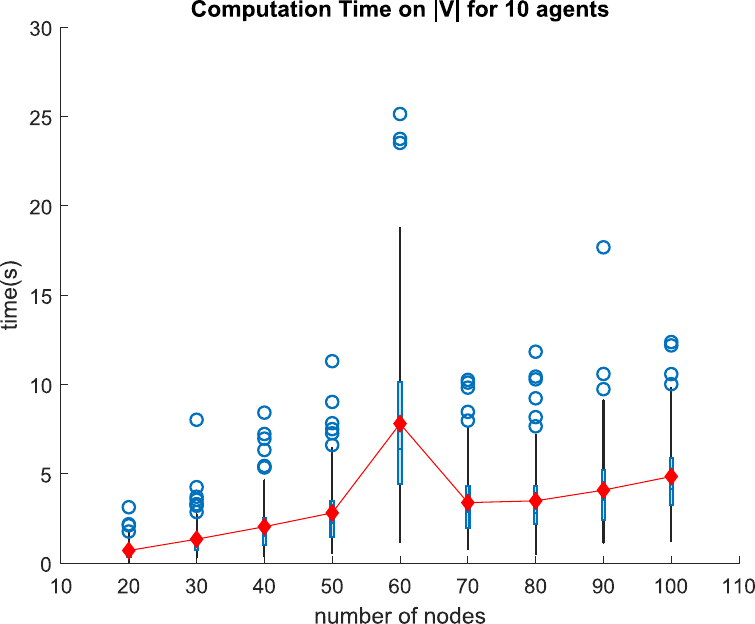}\par
\includegraphics[width=0.33\textwidth]{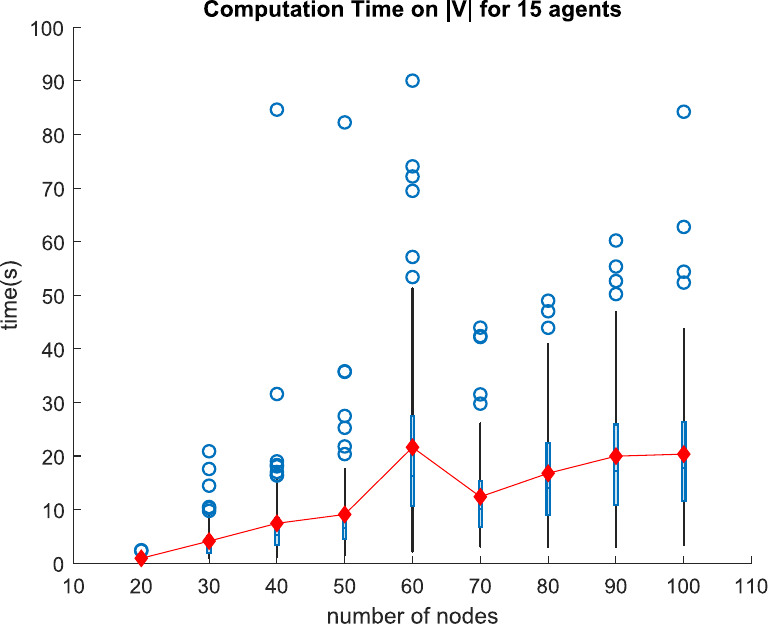}\par
\end{multicols}
\caption{Average Running Time per $|V|$.}
\label{fig:nodes_t_tw}
\end{figure}
In Figures~\ref{fig:nodes_rr_tw} and~\ref{fig:nodes_t_tw} we report the reduction ratio and the running time (in seconds), respectively, wrt $|V|$, for three values of $|P|$. The average is highlighted in red. Figure~\ref{fig:nodes_rr_tw} shows that the reduction ratio decreases wrt the number of nodes. Moreover,  it is much higher in those instances with a higher number of agents $|P|$. This suggests that the local search based on agent distance is not efficient in crowded scenarios, with few unoccupied ndoes.
Figure~\ref{fig:nodes_t_tw} shows that the computation time increases wrt to the number of nodes, for all the number of agents tested.
\begin{figure}[H]
\begin{multicols}{3}
\centering
\includegraphics[width=0.33\textwidth]{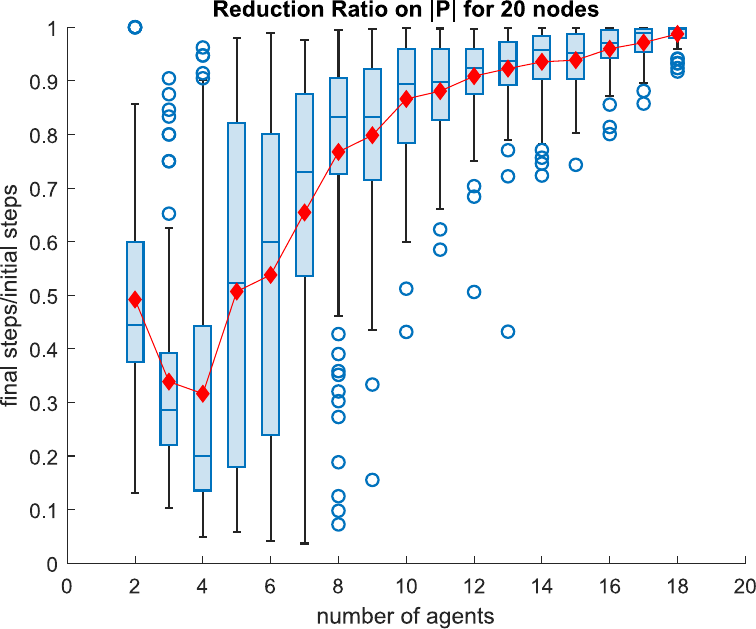}\par
\includegraphics[width=0.33\textwidth]{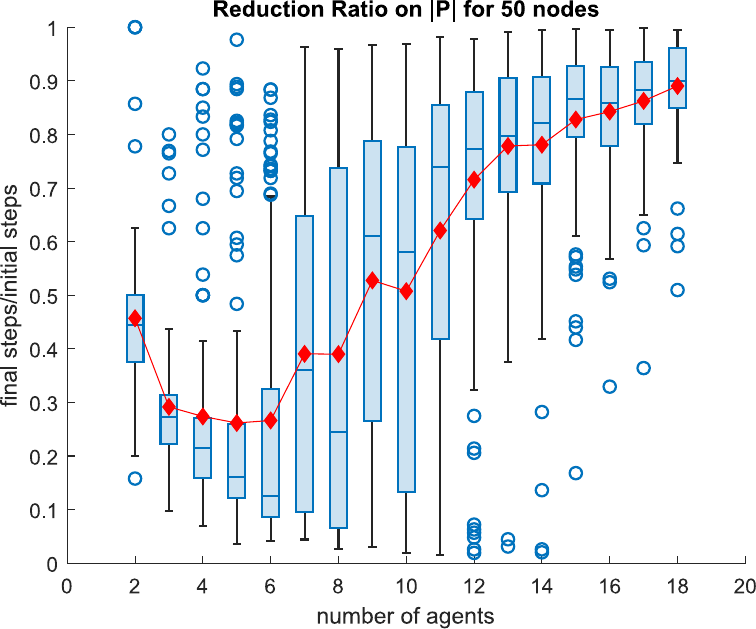}\par
\includegraphics[width=0.33\textwidth]{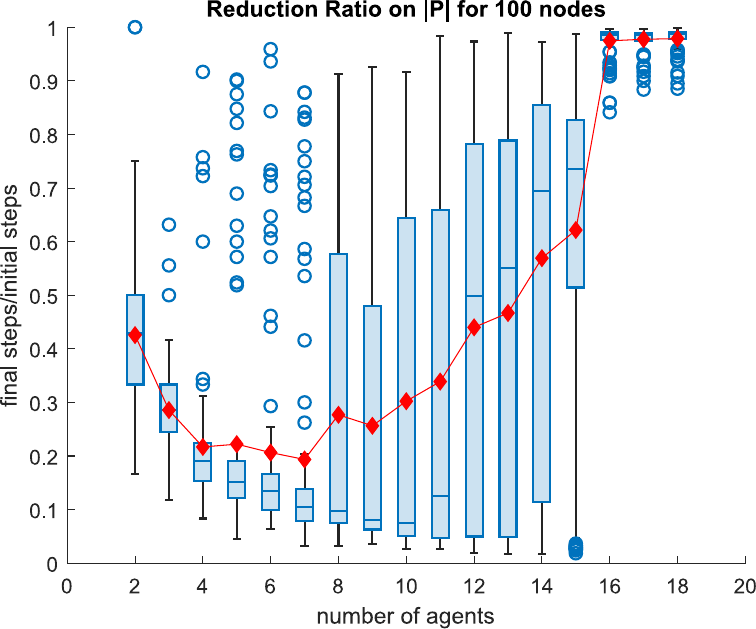}\par
\end{multicols}
\caption{Average Reduction Ratio per $|P|$.}
\label{fig:agents_rr_tw}
\end{figure}
\begin{figure}[H]
\begin{multicols}{3}
\centering
\includegraphics[width=0.33\textwidth]{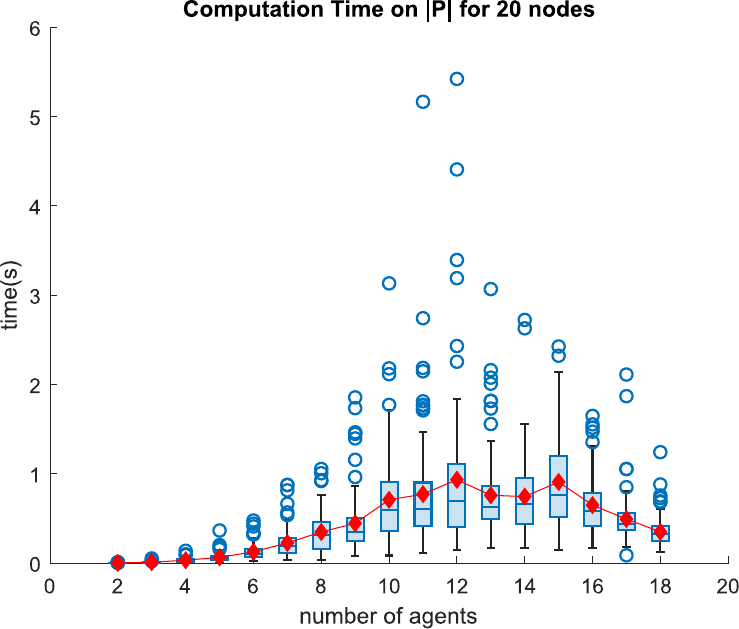}\par
\includegraphics[width=0.33\textwidth]{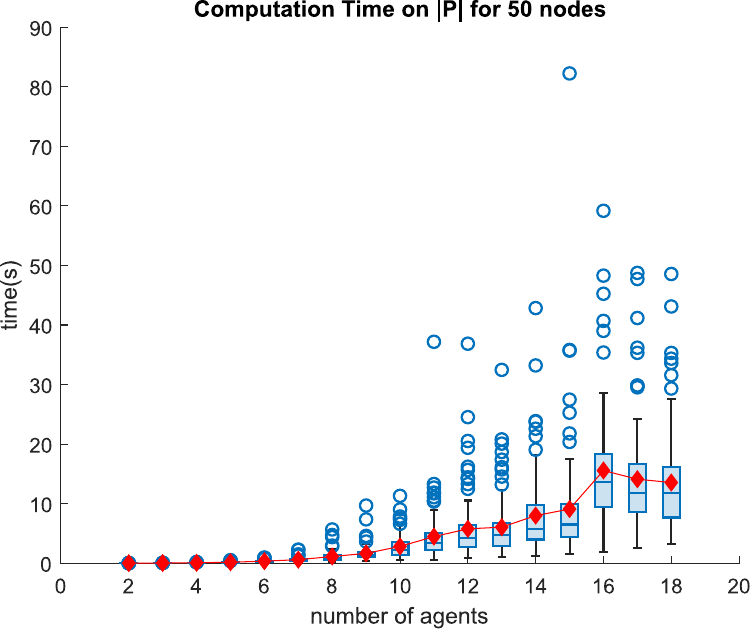}\par
\includegraphics[width=0.33\textwidth]{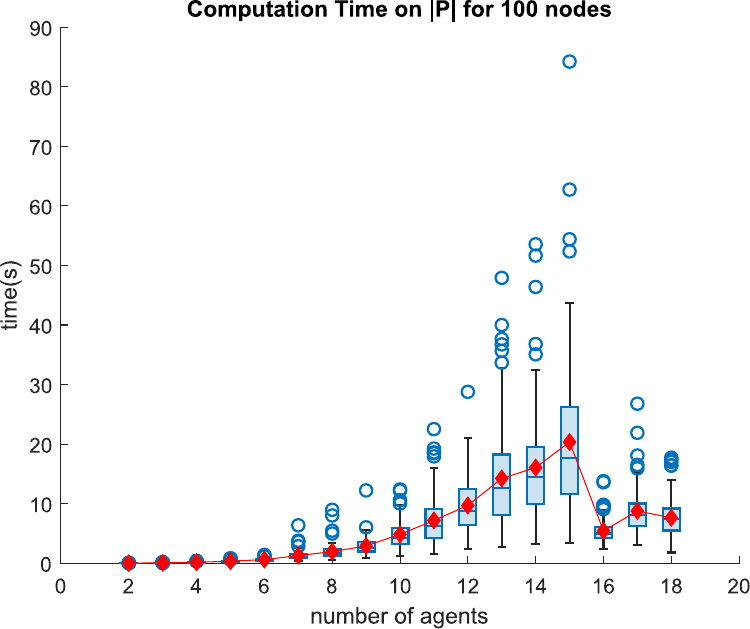}\par
\end{multicols}
\caption{Average Running Time per $|P|$.}
\label{fig:agents_t_tw}
\end{figure}

Figures~\ref{fig:agents_rr_tw} and~\ref{fig:agents_t_tw} represent the reduction ratio and the computation times wrt $|P|$, for three values of $|V|$, with the average highlighted in red. The reduction ratio decreases where the number $|P|$ of agents varies from $2$ to $4-7$ (depending on $|P|$), and then increases wrt $|P|$. The running time increases when $P$ varies from $2$ to $14-16$, and then decreases. Likely, this reduction in computational time depends on the fact that the local search procedure is not effective for larger values of $|P|$ (as suggested by Figure~\ref{fig:agents_rr_tw}), so that the algorithm performs fewer local search iterations. In general, Figures~\ref{fig:nodes_rr_tw} and~\ref{fig:agents_rr_tw} suggest that the local search algorithm with agent distance is more effective in less croded instances, where few agents move inside a large graph.  It is also worthwhile to remark that 
Algorithm~\ref{alg1} tends to display opposite behaviors when run with the path and the agent distance. This suggests the use of the alternated neighborhood search.

\subsection{Alternate Neighborhood Search}
Finally, in the last set of experiments, we tested Algorithm~\ref{alg_alt}.
\begin{figure}[H]
\begin{multicols}{3}
\centering
\includegraphics[width=0.33\textwidth]{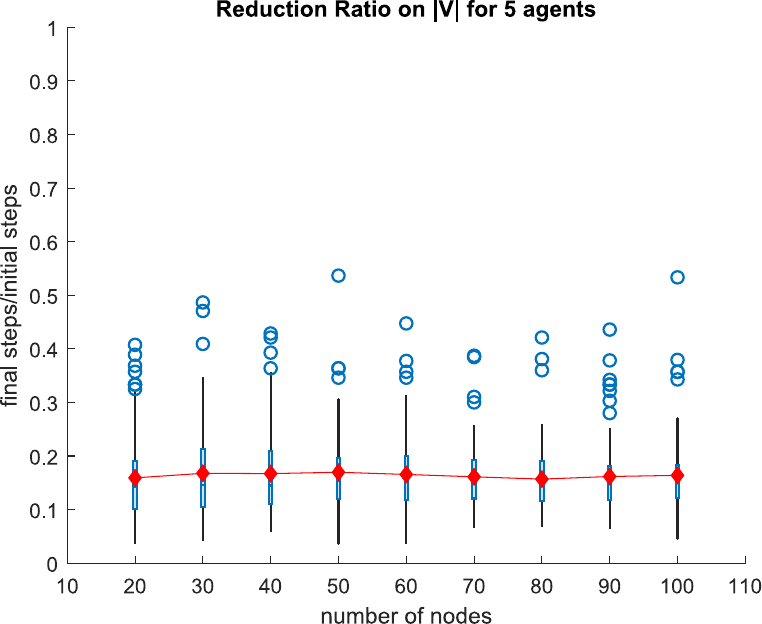}\par
\includegraphics[width=0.33\textwidth]{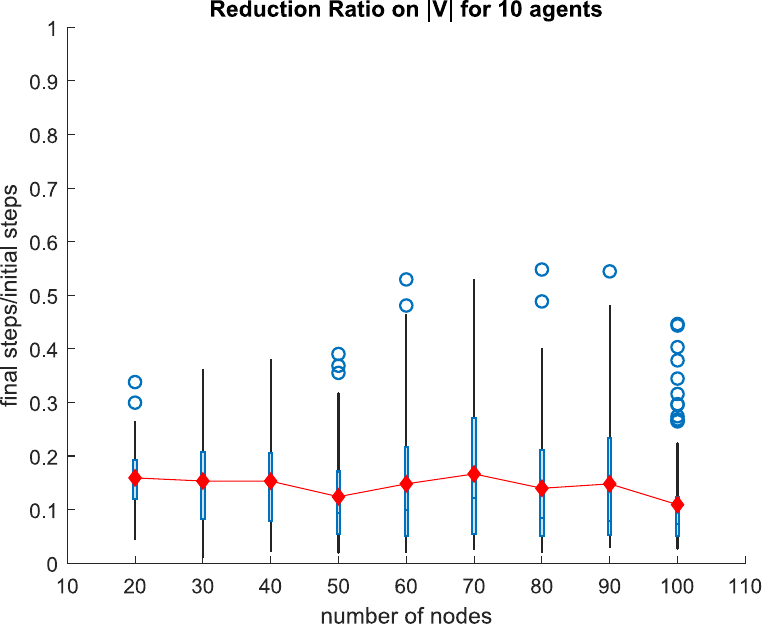}\par
\includegraphics[width=0.33\textwidth]{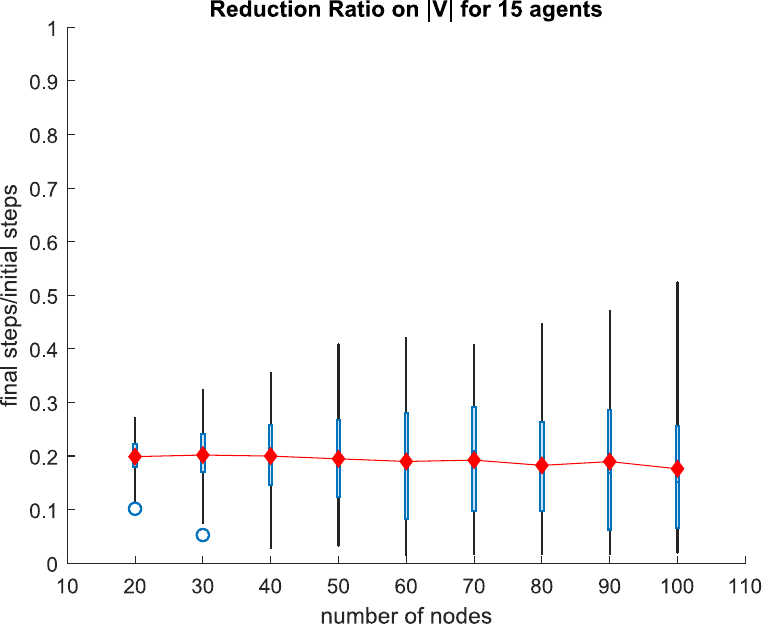}\par
\end{multicols}
\caption{Average Reduction Ratio per $|V|$.}
\label{fig:nodes_rr_both}
\end{figure}
\begin{figure}[H]
\begin{multicols}{3}
\centering
\includegraphics[width=0.33\textwidth]{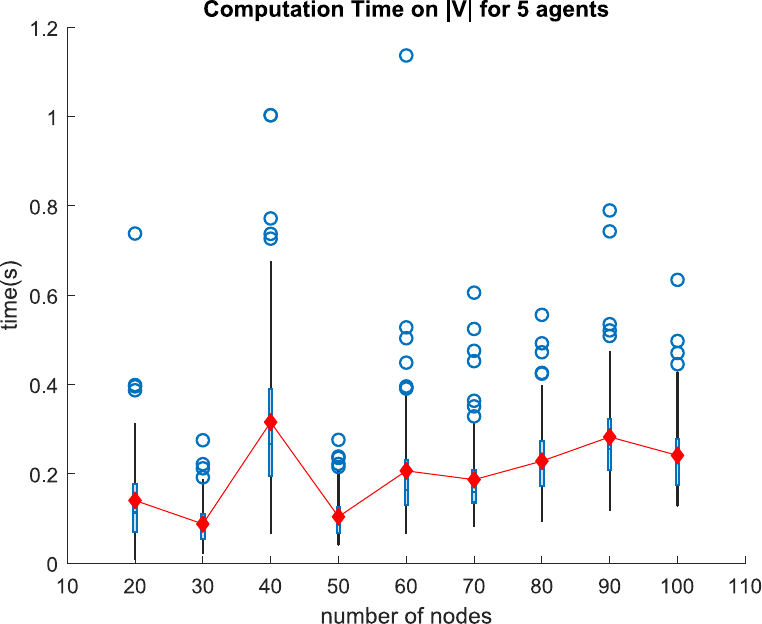}\par
\includegraphics[width=0.33\textwidth]{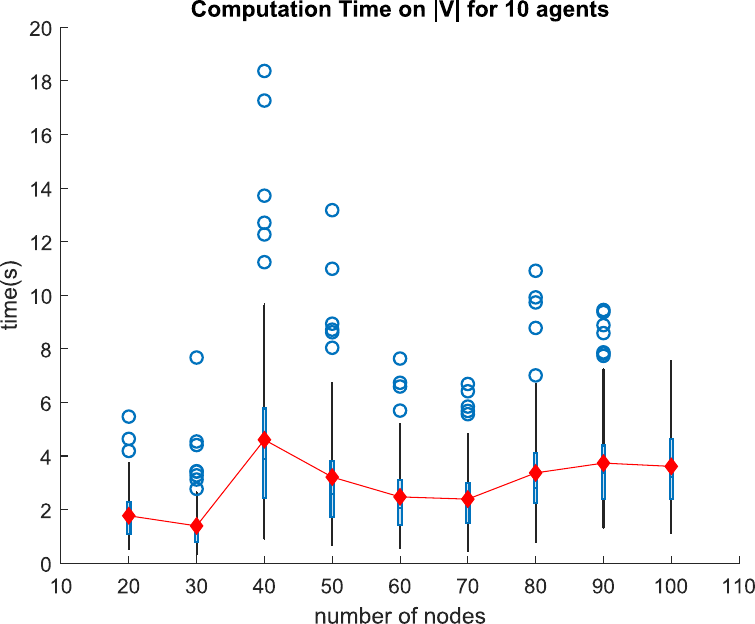}\par
\includegraphics[width=0.33\textwidth]{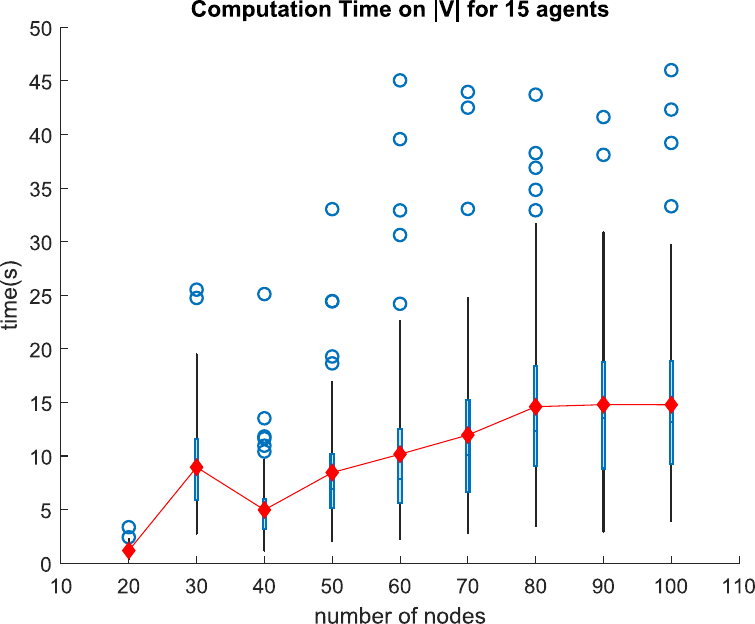}\par
\end{multicols}
\caption{Average Running Time per $|V|$.}
\label{fig:nodes_t_both}
\end{figure}
In Figures \ref{fig:nodes_rr_both} and \ref{fig:nodes_t_both}, we report the reduction ratio and the running time (in seconds), respectively, wrt $|V|$ for three different values of $|P|$. The average is highlighted in red. The results show that the reduction ratio is nearly constant wrt every number of nodes tested, whereas the running time fluctuates and slightly increases.
\begin{figure}[H]
\begin{multicols}{3}
\centering
\includegraphics[width=0.33\textwidth]{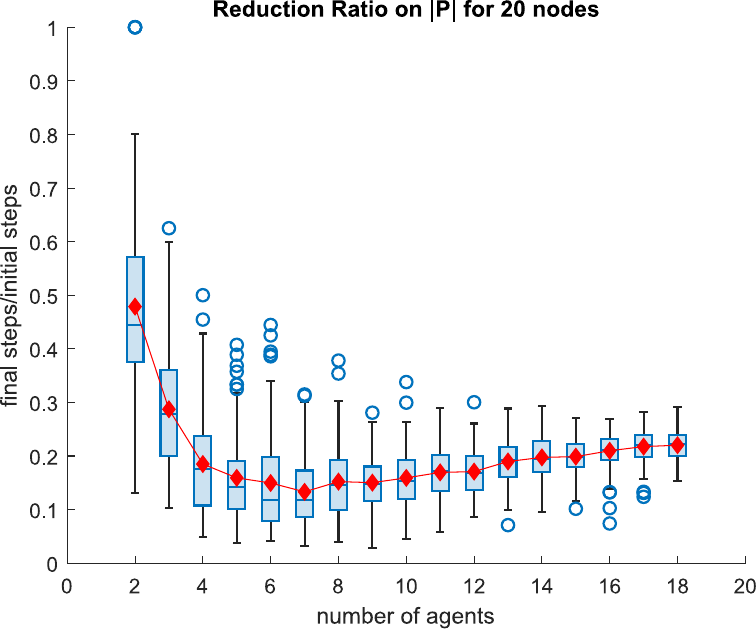}\par
\includegraphics[width=0.33\textwidth]{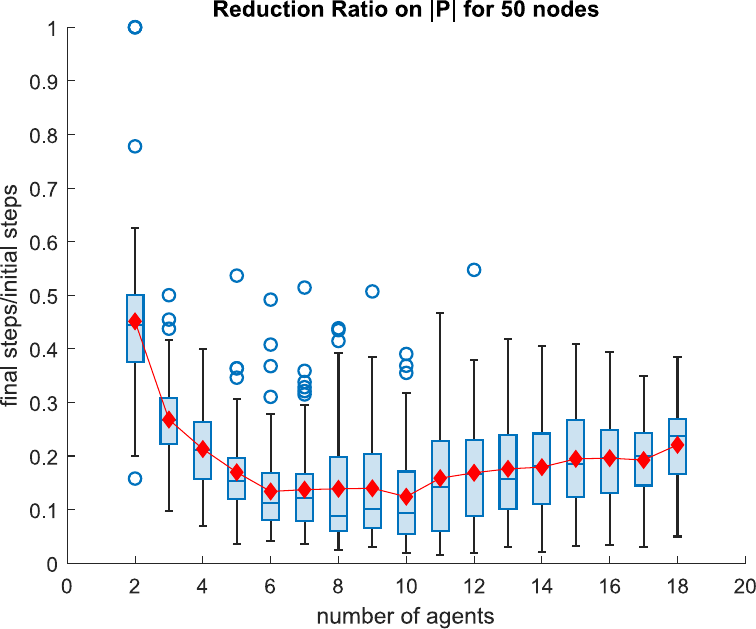}\par
\includegraphics[width=0.33\textwidth]{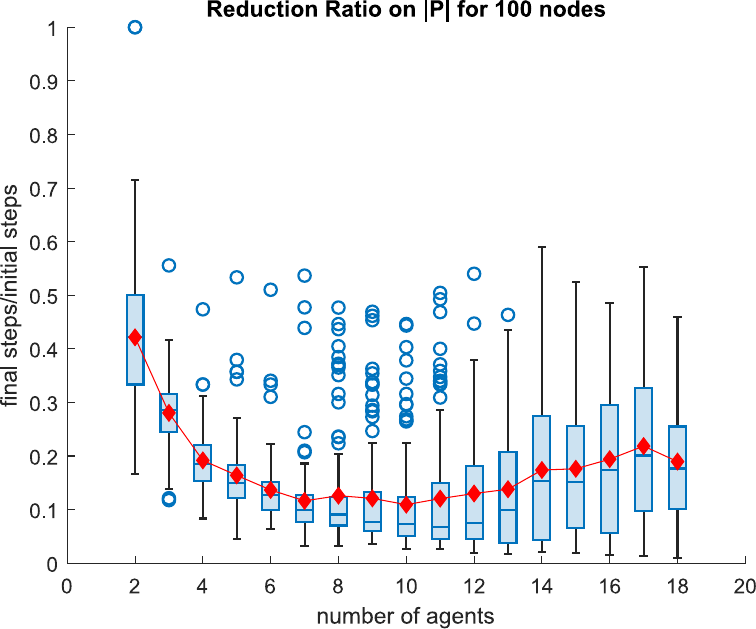}\par
\end{multicols}
\caption{Average Reduction Ratio per $|P|$.}
\label{fig:agents_rr_both}
\end{figure}
\begin{figure}[H]
\begin{multicols}{3}
\centering
\includegraphics[width=0.33\textwidth]{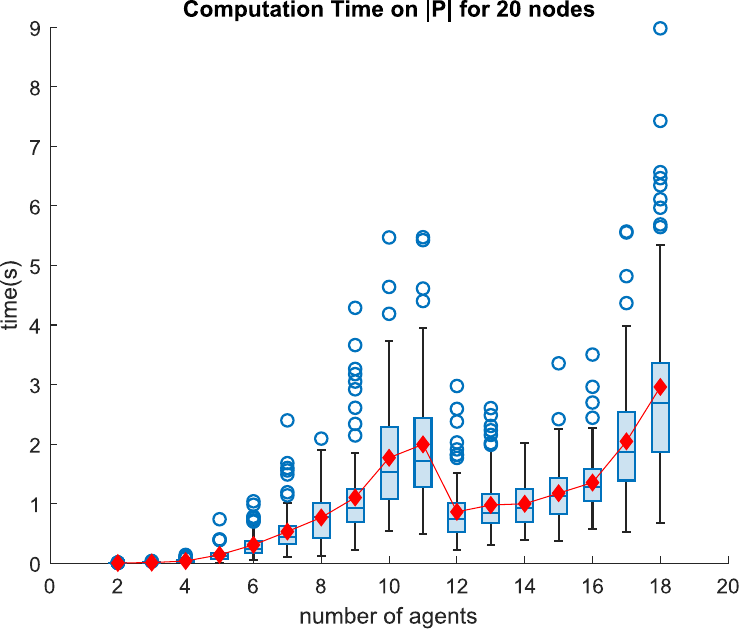}\par
\includegraphics[width=0.33\textwidth]{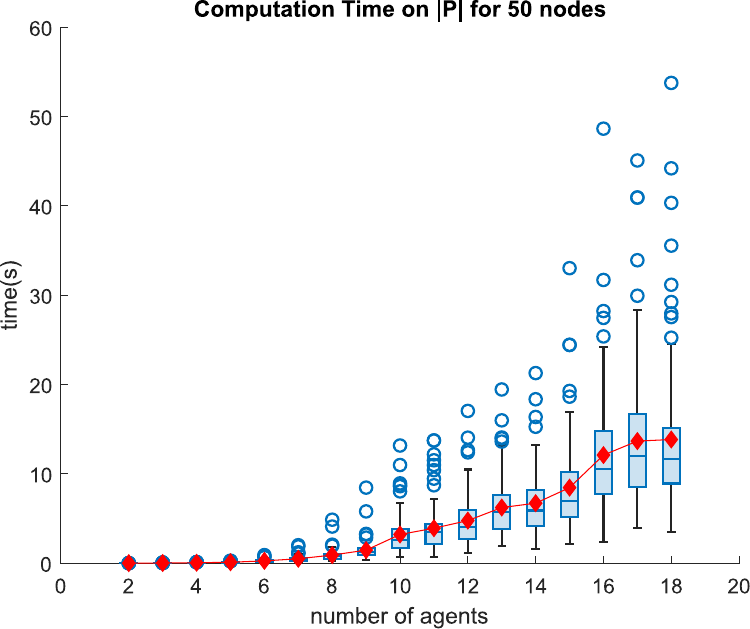}\par
\includegraphics[width=0.33\textwidth]{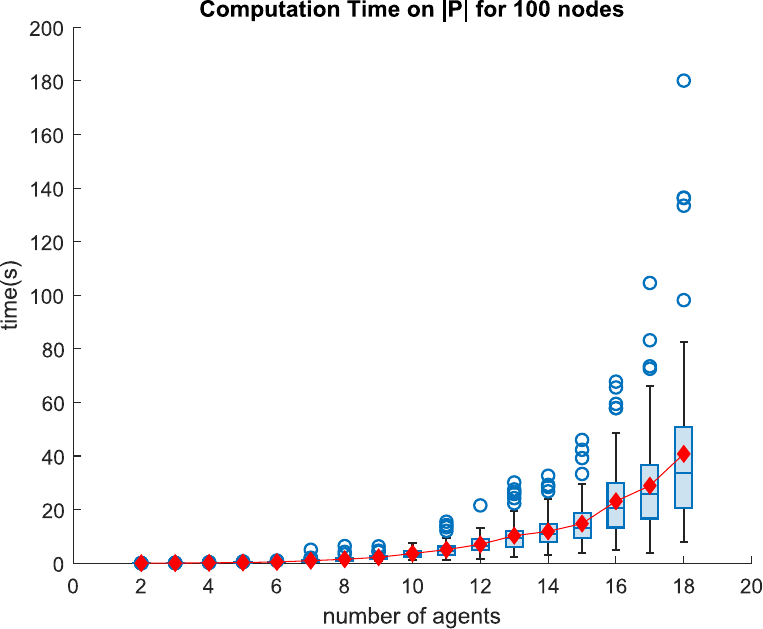}\par
\end{multicols}
\caption{Average Running Time per $|P|$.}
\label{fig:agents_t_both}
\end{figure}
Figures~\ref{fig:agents_rr_both} and~\ref{fig:agents_t_both} show the reduction ratio and the computation times wrt $|P|$, for three different values of $|V|$, with the average highlighted in red. The reduction ratio has a similar behavious for all the three values of $|V|$. It decreases for a lower number of agents, and then increases for larger values of $|P|$. The running time generally increases with the number of agents. This algorithm performs well for nearly all the instances, with the worst results (in terms of reduction ratio) for instances with a lower number of agents. Likely, this is because these instances are simpler, and the initial plans are closer to the local optimum, in comparision to the other instances.
\newline\newline\noindent
We can now compare the average results of all the three set of experiments. 
\begin{figure}[H]
\begin{multicols}{3}
\centering
\includegraphics[width=0.33\textwidth]{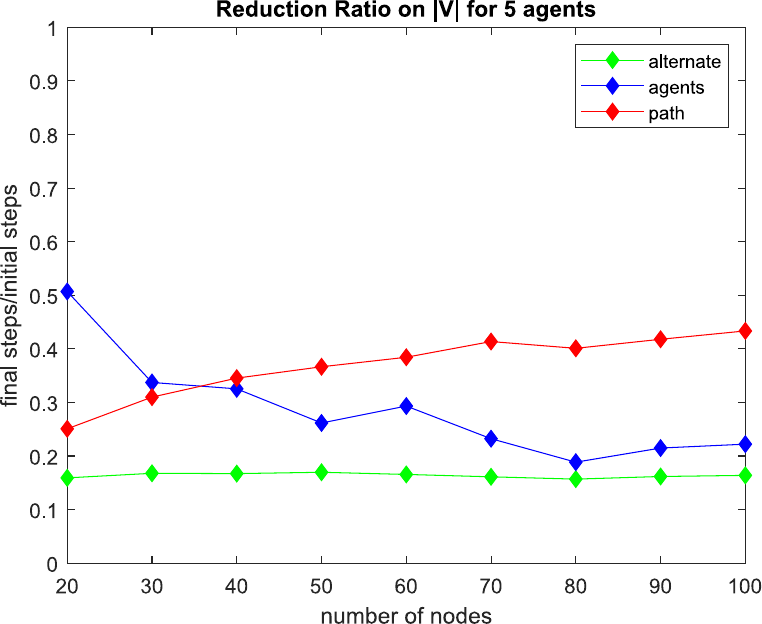}\par
\includegraphics[width=0.33\textwidth]{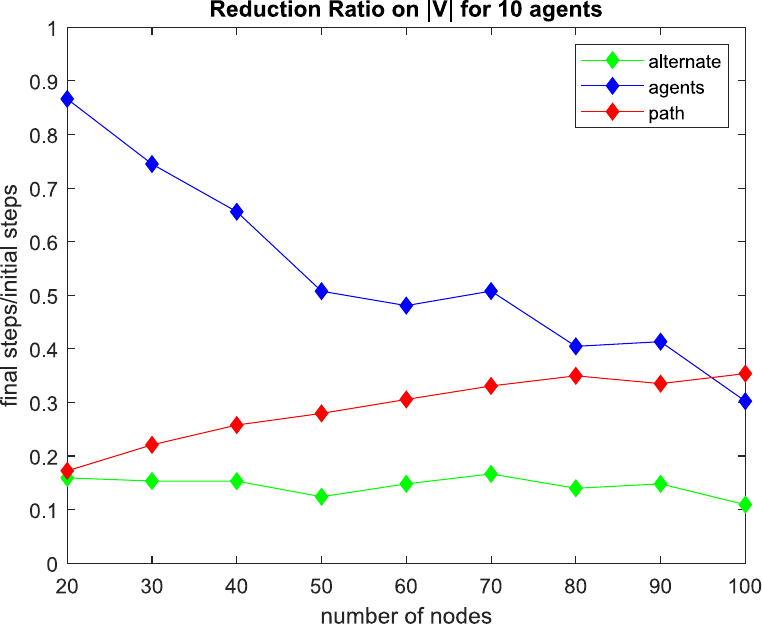}\par
\includegraphics[width=0.33\textwidth]{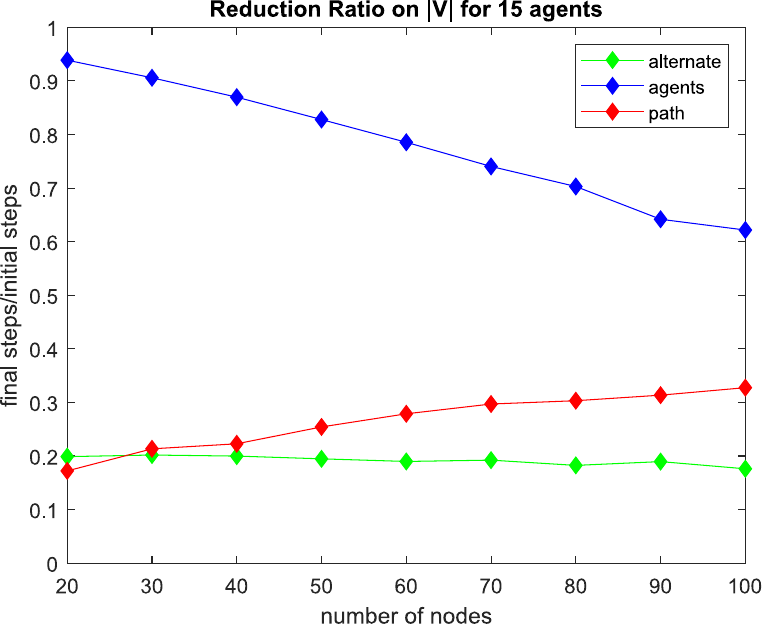}\par
\end{multicols}
\caption{Average Reduction Ratio per $|V|$.}
\label{fig:nodes_rr}
\end{figure}
\begin{figure}[H]
\begin{multicols}{3}
\centering
\includegraphics[width=0.33\textwidth]{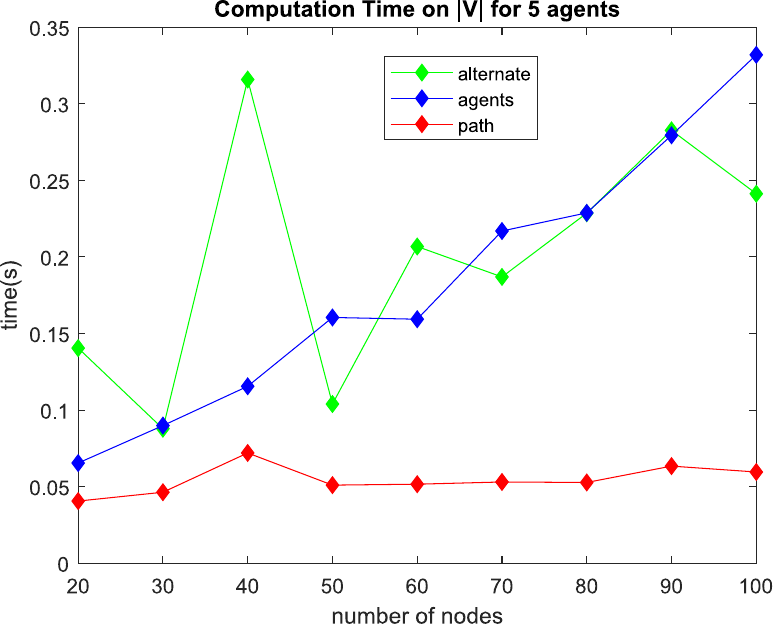}\par
\includegraphics[width=0.33\textwidth]{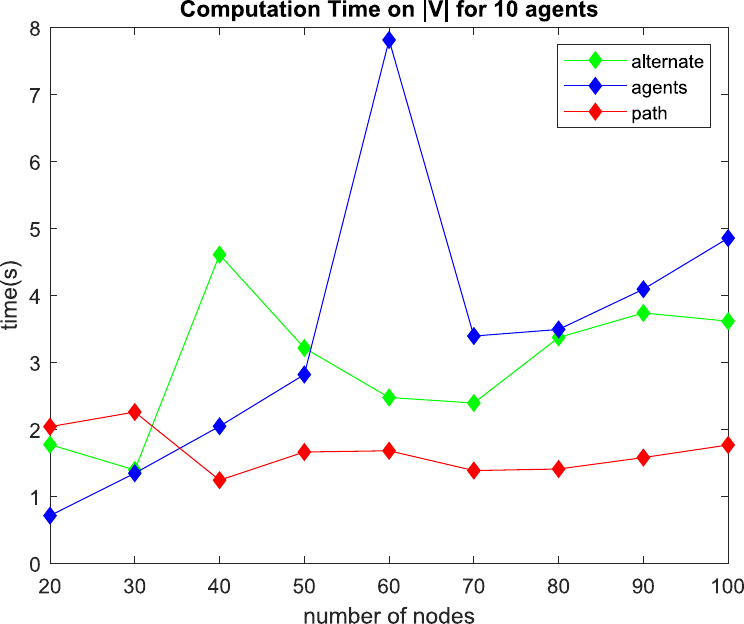}\par
\includegraphics[width=0.33\textwidth]{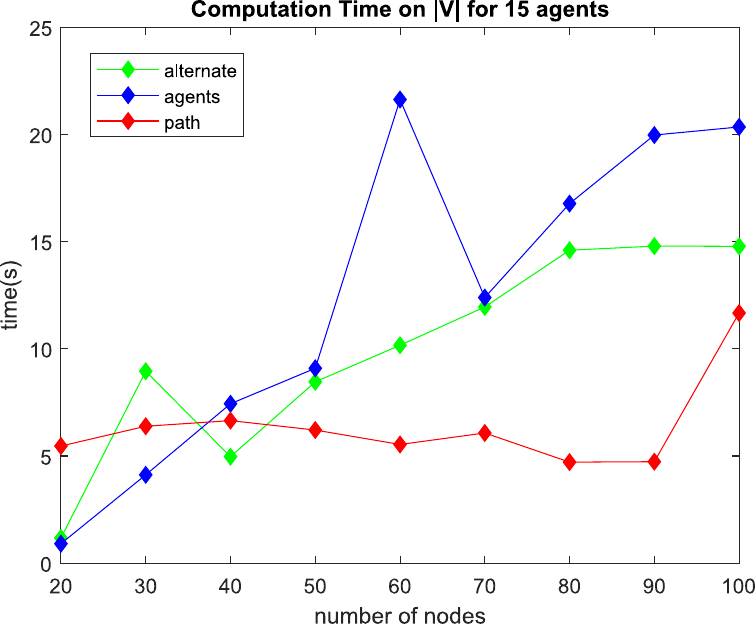}\par
\end{multicols}
\caption{Average Running Time per $|V|$.}
\label{fig:nodes_t}
\end{figure}

Figures~\ref{fig:nodes_rr} and~\ref{fig:nodes_t} compare the reduction rates and the computation times of all approaches, wrt the number of nodes $|V|$. The Alternate Neighborhood Search always returns the shortest plan, in comparison to the other approaches. It is interesting to compare the results of the u-agents distance and the path distance. In the first plot of Figure~\ref{fig:nodes_rr}, obtained using a low number of agents ($|P|=5$), the u-agents distance has better results for $|V|>30$. On the other hand, for a high number of agents, such as the case of the third plot ($|P|=15$), the path distance approach returns much shorter plans for all the number of nodes. This behavior is related to the fact that the u-agents distance approach is not as effective 
as the path distance approach in crowded graphs (i.e., where the number of agents is high compared to the number of nodes). Figure~\ref{fig:nodes_t} shows that the algorithm that uses the path-distance is overall faster than the other two algorithms, whereas the other two algorithms have comparable running times that increase wrt the number of nodes.

\begin{figure}[H]
\begin{multicols}{3}
\centering
\includegraphics[width=0.33\textwidth]{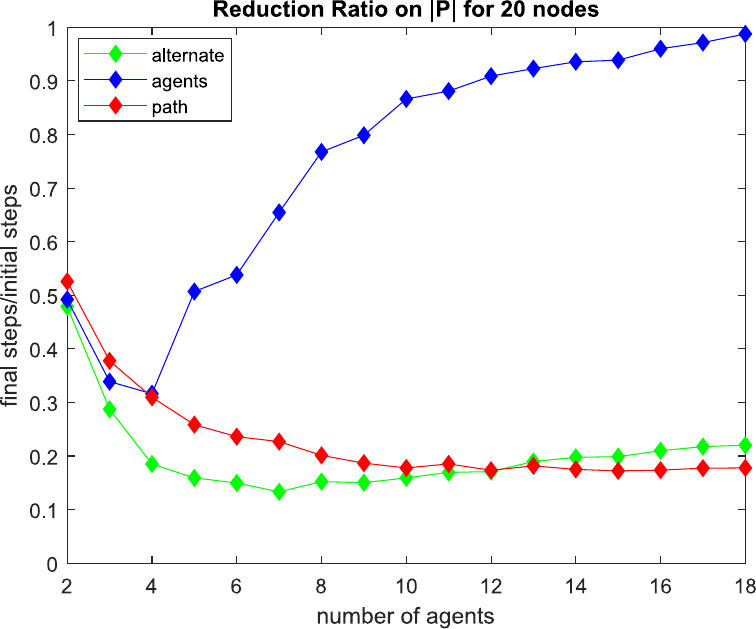}\par
\includegraphics[width=0.33\textwidth]{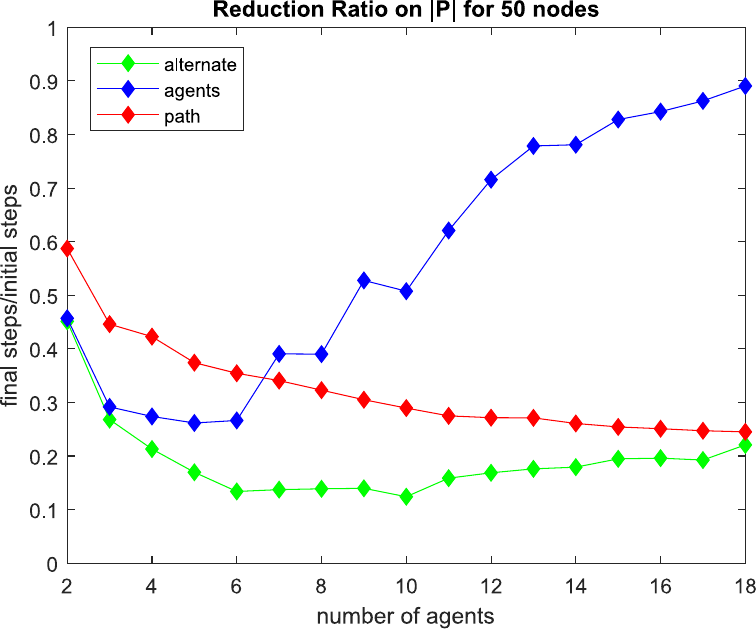}\par
\includegraphics[width=0.33\textwidth]{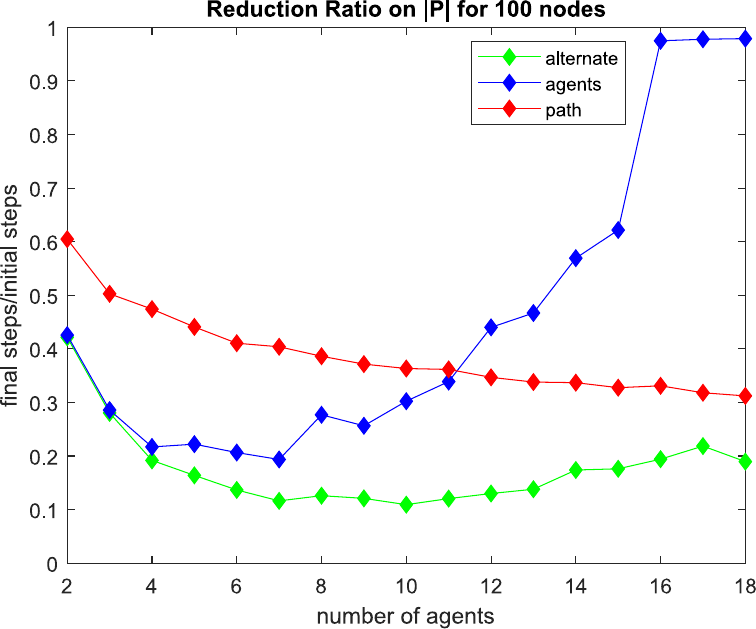}\par
\end{multicols}
\caption{Average Reduction Ratio per $|P|$.}
\label{fig:agents_rr}
\end{figure}
\begin{figure}[H]
\begin{multicols}{3}
\centering
\includegraphics[width=0.33\textwidth]{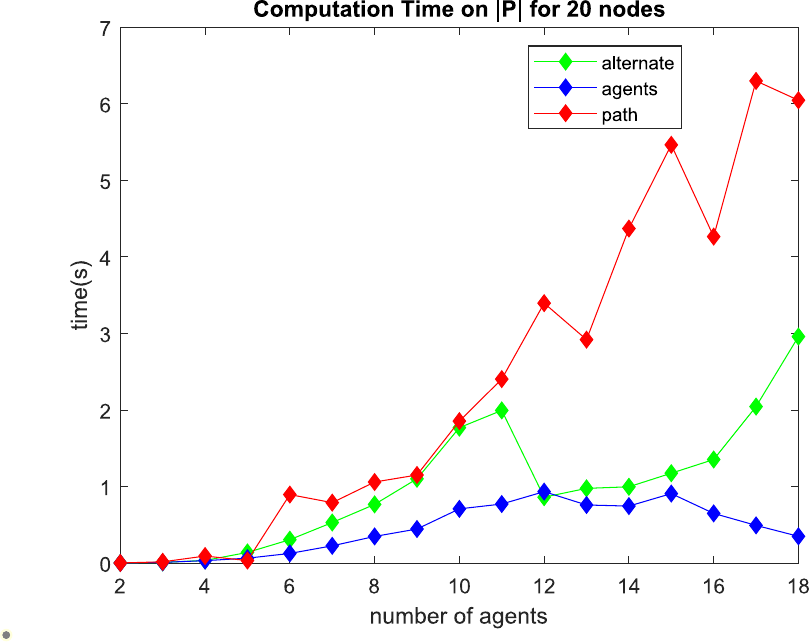}\par
\includegraphics[width=0.33\textwidth]{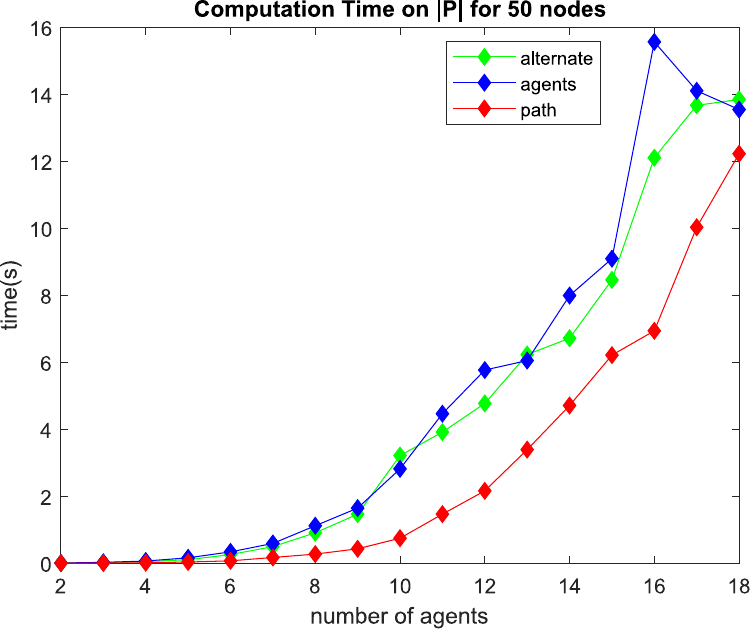}\par
\includegraphics[width=0.33\textwidth]{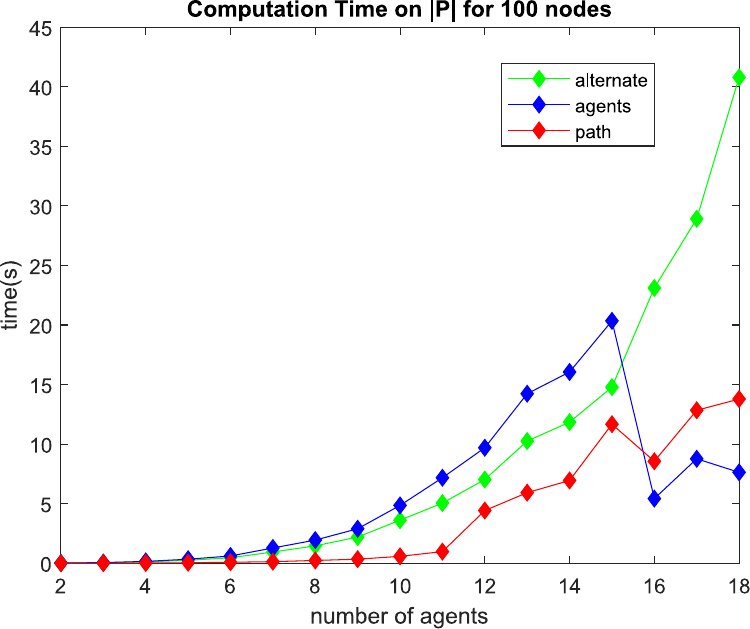}\par
\end{multicols}
\caption{Average Running Time per $|P|$.}
\label{fig:agents_t}
\end{figure}

Figures~\ref{fig:agents_rr} and~\ref{fig:agents_t} compare the reduction rates and the computation times of all the approaches, wrt the number of agents $|P|$. As already observed for the results wrt $|V|$, the performances of the three approaches depending on both $V$ and $|P|$. In the plots of Figure~\ref{fig:agents_rr}, the reduction rates of both Algorithm~\ref{alg1} with sum-min distance and Algorithm~\ref{alg_alt} decrease wrt the number of agents, whereas the reduction rate of Algorithm~\ref{alg1} with u-agents distance first slightly decreases and then increases, reaching values near $1$ (that corresponds to a negligible reduction of the length of the starting plan). This increase is much faster in the first plot ($|V|=20$), and slows down for higher numbers of nodes. In particular, in the last plot ($|V|=100$), for $|P|<12$ the algorithm with u-agents distance presents a lower reduction ratio than the algorithm with sum-min distance. The running time appears to increase polynomially wrt $|P|$ for all the algorithms and all the three values of $|V|$. In the first plot of Figure~\ref{fig:agents_t}, the three sets of experiments have different behaviors. The running time of the algorithm that uses the sum-min distance is much higher than the others. This is probably due to a larger number of iteration with respect to the algorithm that uses the u-agents distance. 

\section{Conclusion and future works}

We propose a method to improve feasible solutions to MAPF problems. To obtain shorter solutions, we search the solution space using a \textit{Dynamic Programming} algorithm. To reduce the number of states explored, and consequently, the time complexity, we iteratively define a neighborhood of the given solution in which to search for better ones. The neighborhoods can be defined using different kinds of distances. In this paper, we focused on two distances: the first one is defined based on the length of the shortest path between nodes, and the second one considers the number of agents that vary their paths between two solutions. In all cases, the proposed algorithms have polynomial time complexity with respect to the number of agents and the number of nodes. We tested this approach using each distance separately and together. The experiments show that using both distances, alternating the neighborhoods where the solution is searched, is a valid procedure to find solutions close to the optimal ones.
\newpage 
\appendix
\bibliographystyle{plain}
\bibliography{bibliography}

\begin{thebibliography}{10}

\bibitem{alotaibi2016}
Ebtehal Turki~Saho Alotaibi and Hisham Al-Rawi.
\newblock Push and spin: {A} complete multi-robot path planning algorithm.
\newblock In {\em 2016 14th {International} {Conference} on {Control},
  {Automation}, {Robotics} and {Vision} ({ICARCV})}, pages 1--8, Phuket,
  Thailand, November 2016. IEEE.

\bibitem{ardizzoni2022}
S.~Ardizzoni, I.~Saccani, L.~Consolini, and M.~Locatelli.
\newblock Multi-{Agent} {Path} {Finding} on {Strongly} {Connected} {Digraphs}.
\newblock In {\em 2022 {IEEE} 61st {Conference} on {Decision} and {Control}
  ({CDC})}, pages 7194--7199, Cancun, Mexico, December 2022. IEEE.

\bibitem{ardizzoni231}
S.~Ardizzoni, I.~Saccani, L.~Consolini, and M.~Locatelli.
\newblock Local optimization of {MAPF} solutions on directed graphs.
\newblock In {\em 2023 62nd IEEE Conference on Decision and Control (CDC)},
  pages 8081--8086, 2023.

\bibitem{ardizzoni2024}
Stefano Ardizzoni, Irene Saccani, Luca Consolini, Marco Locatelli, and Bernhard
  Nebel.
\newblock An algorithm with improved complexity for pebble motion/multi-agent
  path finding on trees.
\newblock {\em Journal of Artificial Intelligence Research}, 79:483--514, 2024.

\bibitem{botea2018}
Adi Botea, Davide Bonusi, and Pavel Surynek.
\newblock Solving {Multi}-agent {Path} {Finding} on {Strongly} {Biconnected}
  {Digraphs}.
\newblock {\em Journal of Artificial Intelligence Research}, 62:273--314, June
  2018.

\bibitem{goraly2010}
Gilad Goraly and Refael Hassin.
\newblock Multi-{Color} {Pebble} {Motion} on {Graphs}.
\newblock {\em Algorithmica}, 58(3):610--636, November 2010.

\bibitem{khorshid2021}
Mokhtar Khorshid, Robert Holte, and Nathan Sturtevant.
\newblock A {Polynomial}-{Time} {Algorithm} for {Non}-{Optimal} {Multi}-{Agent}
  {Pathfinding}.
\newblock {\em Proceedings of the International Symposium on Combinatorial
  Search}, 2(1):76--83, August 2021.

\bibitem{kornhauser1984}
D.~Kornhauser, G.~Miller, and P.~Spirakis.
\newblock Coordinating {Pebble} {Motion} {On} {Graphs}, {The} {Diameter} {Of}
  {Permutation} {Groups}, {And} {Applications}.
\newblock In {\em 25th {Annual} {Symposium} {on Foundations} of {Computer}
  {Science}, 1984.}, pages 241--250, Singer Island, FL, 1984. IEEE.

\bibitem{krontiris2021}
Athanasios Krontiris, Ryan Luna, and Kostas Bekris.
\newblock From {Feasibility} {Tests} to {Path} {Planners} for {Multi}-{Agent}
  {Pathfinding}.
\newblock {\em Proceedings of the International Symposium on Combinatorial
  Search}, 4(1):114--122, August 2021.

\bibitem{li2021}
Jiaoyang Li, Zhe Chen, Daniel Harabor, Peter~J Stuckey, and Sven Koenig.
\newblock Anytime multi-agent path finding via large neighborhood search.
\newblock In {\em International Joint Conference on Artificial Intelligence
  2021}, pages 4127--4135. Association for the Advancement of Artificial
  Intelligence (AAAI), 2021.

\bibitem{li2022}
Jiaoyang Li, Zhe Chen, Daniel Harabor, Peter~J Stuckey, and Sven Koenig.
\newblock Mapf-lns2: Fast repairing for multi-agent path finding via large
  neighborhood search.
\newblock In {\em Proceedings of the AAAI Conference on Artificial
  Intelligence}, volume~36, pages 10256--10265, 2022.

\bibitem{pisinger2019}
David Pisinger and Stefan Ropke.
\newblock Large neighborhood search.
\newblock {\em Handbook of metaheuristics}, pages 99--127, 2019.

\bibitem{sharon2015}
Guni Sharon, Roni Stern, Ariel Felner, and Nathan~R. Sturtevant.
\newblock Conflict-based search for optimal multi-agent pathfinding.
\newblock {\em Artificial Intelligence}, 219:40--66, February 2015.

\bibitem{shaw1998}
Paul Shaw.
\newblock Using constraint programming and local search methods to solve
  vehicle routing problems.
\newblock In {\em International conference on principles and practice of
  constraint programming}, pages 417--431. Springer, 1998.

\bibitem{silver2021}
David Silver.
\newblock Cooperative {Pathfinding}.
\newblock {\em Proceedings of the AAAI Conference on Artificial Intelligence
  and Interactive Digital Entertainment}, 1(1):117--122, September 2021.

\bibitem{wilde2014}
B.~de Wilde, A.~W. Ter~Mors, and C.~Witteveen.
\newblock Push and {Rotate}: a {Complete} {Multi}-agent {Pathfinding}
  {Algorithm}.
\newblock {\em Journal of Artificial Intelligence Research}, 51:443--492,
  October 2014.

\bibitem{yu2013}
Jingjin Yu and Steven LaValle.
\newblock Structure and intractability of optimal multi-robot path planning on
  graphs.
\newblock In {\em Proceedings of the AAAI Conference on Artificial
  Intelligence}, volume~27, pages 1443--1449, 2013.

\end{thebibliography}
\end{document}